\documentclass[aps,prl,twocolumn,superscriptaddress]{revtex4}
\usepackage{mathrsfs}
\usepackage{epsfig}
\usepackage{graphicx}
\usepackage{amsfonts}
\usepackage[figuresright]{rotating}
\usepackage{amssymb}
\usepackage{amsmath}
\usepackage{dcolumn}
\usepackage{bm}
\usepackage{color}

\def\be{\begin{equation}} \def\ee{\end{equation}}
\def\bea{\begin{eqnarray}} \def\eea{\end{eqnarray}}

\newcommand{\WQCASQC} {Wilczek Quantum Center and Key Laboratory of Artificial Structures and Quantum Control, School of Physics and Astronomy, Shanghai Jiao Tong University, Shanghai 200240, China}

\newcommand{\SRCQC}{Shanghai Research Center for Quantum Sciences, Shanghai 201315, China}

\begin{document}
\title{Thermal melting of discrete time crystals: a dynamical phase transition induced by thermal fluctuations}

\author{Mingxi Yue}
\thanks{These authors contributed equally to this work.}
\affiliation{\WQCASQC}

\author{Xiaoqin Yang}
\thanks{These authors contributed equally to this work.}
\affiliation{\WQCASQC}

\author{Zi Cai}
\email{zcai@sjtu.edu.cn}
\affiliation{\WQCASQC}
\affiliation{\SRCQC}

\begin{abstract}  

The stability of a discrete  time crystal against thermal fluctuations has been studied numerically  by solving a stochastic Landau-Lifshitz-Gilbert equation of a periodically-driven classical system composed of interacting spins, each of which couples to a thermal bath. It is shown that in the thermodynamic limit, even though the long-range temporal crystalline order is stable at low temperature, it is melting above a critical temperature, at which the system experiences a non-equilibrium phase transition. The critical behaviors of the continuous phase transition have been systematically investigated, and it is shown that  despite the genuine non-equilibrium feature of such a periodically driven system, its critical properties fall into the 3D Ising universality class with a dynamical exponent ($z=2$) identical to that in the critical  dynamics of kinetic Ising model without driving.



\end{abstract}


\maketitle

{\it Introduction --}
Spontaneous symmetry breakings (SSB) and universality classes are among the most fundamental concepts in  modern physics: the former is used to characterize different phases of matter while the latter is to categorize the transitions between them. Compared to equilibrium systems, the physics in out-of-equilibrium systems is much richer but less known in general. The evolutionary processes give rise to increasing richness of the paradigm of spontaneous symmetry breaking and universality classes, which could take place not only in space but also in time,  the latter opens up new opportunities to explore novel states of matter (the time crystal(TC)\cite{Wilczek2012} for instance) and dynamical universality classes ( the critical dynamical phenomena\cite{Hohenberg1977,Taeuber2014} or the Kardar-Parisi-Zhang (KPZ) universality class\cite{Kardar1986} for instance)  beyond the scope of equilibrium physics.

As a prototypical example of spontaneous symmetry breaking in time domain, the time crystal phase\cite{Wilczek2012}, in its different forms\cite{Shapere2012,Li2012,Wilczek2013,Sacha2015,Else2016,Khemani2016,Yao2017,Russomanno2017,Gong2018,Huang2018,Iemini2018,Das2018,Zhu2019,Liao2019,Kozin2019,Cai2020,Chinzei2020,Lyu2020,Choudhury2021}, has attracted considerable interests in past decades. In spite of being proven absent in thermal equilibrium\cite{Bruno2013,Watanabe2015}, such an intriguing phase has been observed in periodically driven non-equilibrium settings\cite{Choi2017,Zhang2017,Autti2018,Trager2021,Kessler2021,Mi2021,Frey2021}. These systems exhibit oscillations with period doubling with respect to that of the external driving, thus spontaneously break the discrete time translational symmetry(TTS) from symmetry group $\mathbb{Z}$ to $2\mathbb{Z}$. As a phase of matter, a profound question is its stability against perturbations, especially the thermal fluctuations  inevitable in almost all the realistic experimental setups.  From a theoretical point of view, this problem can be considered as a non-equilibrium analogue of thermal melting of the crystalline order, and is of fundamental interest due to its relevance to broader questions of the robustness of the temporal order against fluctuations, as well as the dynamical universality classes in non-equilibrium matter.

Recently,  Yao {\it et al} explored this question by studying a classical Hamiltonian dynamics coupled to a finite-temperature bath, and found an activated discrete time crystal(DTC) whose crystalline order survives to long, but not infinite times at low temperature~\cite{Yao2020}. A question naturally raised is whether there exists DTC phases with true long-range crystalline order persisting forever~\cite{Zhuang2021}? Is it possible for the thermal fluctuation to melt such a dynamical crystalline order, just as it does for conventional crystals?  If so, how to characterize such a  non-equilibrium phase transition and what's the corresponding universality class?

In this paper, we attempt to answer these questions, focusing for simplicity on a three-dimensional (3D) classical periodically-driven interacting spin model, which could exhibit  period doubling with respect to the driving (DTC phase) at zero temperature. The thermal fluctuations are introduced by coupling each spin to a heat bath, which provides both dissipation and noise.  The problem is approached by solving a stochastic  equation of motion (EOM),  where the dynamics of each spin  is governed by a  Landau-Lifshitz-Gilbert (LLG) equation\cite{Lakshmanan2010} augmented by a random  thermal force\cite{William1963}. It has been shown that at high temperature, the DTC order parameter decays exponentially with time, indicating a finite life time of the DTC phase as in the 1D case (activated DTC). With decreasing temperature, the non-equilibrium system experiences a dynamical phase transition from an activated DTC  to a DTC phase with true long-range crystalline order.   It is interesting to show that despite the genuine non-equilibrium feature of such periodically driven systems, its critical properties  are characterized by an Ising universality class with a dynamical exponent ($z=2$) identical with the value in the critical dynamics of undriven systems.



 \begin{figure}[htb]
\includegraphics[width=0.99\linewidth,bb=18 22 525 406]{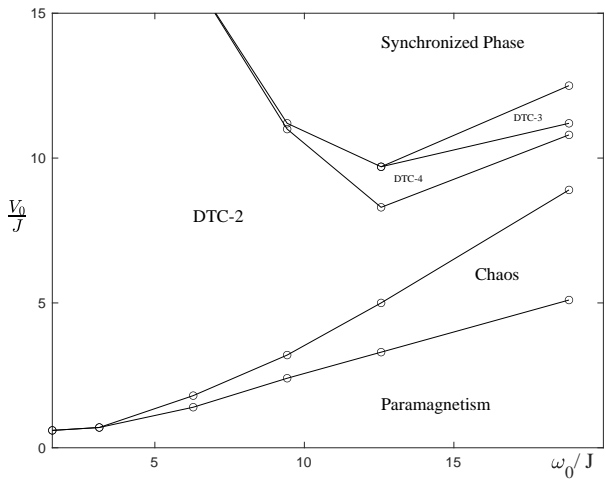}
\caption{(Color online) Zero temperature phase diagram of the long-time behavior of the classical EOM.(\ref{eq:EOM}) with $h_z=\lambda=J$ and $\mathcal{D}=0$. } \label{fig:fig1}
\end{figure}

{\it Model and method --}
The time crystal phases in classical many-body systems have attracted considerable interest recently\cite{Lupo2019,Yao2020,Gambetta2019,Khasseh2019,Hurtado2020,Ye2021,Pizzi2021}. We start with a 3D  classical spin model (transverse Ising model) with a time-dependent coupling strength.  The system Hamiltonian reads:
\begin{equation}
H_s(t)=-V(t)\sum_{\langle ij\rangle} s_i^x s_j^x+\sum_i h_z s_i^z \label{eq:ham}
\end{equation}
where the dynamical variable $\mathbf{s}_i$ is a classical vector with fixed length $|\mathbf{s}_i|=1$.  The summation $\langle ij\rangle$ is over the nearest-neighboring (NN) sites of site $i$ in the cubic lattice with length $L$.   The strength of the interaction $V(t)$  periodically oscillates as $V(t)=V_0\cos\omega_0 t$. $h_z$ is the strength of a uniform magnetic field along z-direction. Throughout this paper we fix  $h_z=J$.  In the absence of thermal bath, the dynamics of each spin can be described by the EOM: $\dot{\mathbf{s}}_i=\mathbf{h}^0_i\times \mathbf{s}_i$, where the effective magnetic field $\mathbf{h}^0_i=[V(t) \bar{s}_i^x, 0, h_z]$ with $\bar{s}_i^x= \frac 12\sum_{\langle j\rangle} s_j^x$  (the summation is over the NN sites of site i).


Even though our driven spin system is a genuine non-equilibrium system whose temperature is ill-defined, usually the degrees freedom of  bath are much larger than those of the system thus the back action of the system to the bath can be neglected, which allows us to consider a thermal bath with a well-defined temperature (T) that doesn't change during the dynamical process.  The effect of the thermal bath can be modeled by methods familiar in the theory of the Brownian motion and other  stochastic processes, where the EOM of each spin is described by a stochastic LLG equation\cite{Lakshmanan2010} as:
\begin{equation}
\dot{\mathbf{s}}_i=\mathbf{h}_i\times \mathbf{s}_i- \lambda  \mathbf{s}_i\times (\mathbf{s}_i\times \mathbf{h}_i) \label{eq:EOM}
\end{equation}
where $\lambda$ is the strength of the friction provided by the local thermal bath, which is fixed as $\lambda=J$ throughout this paper.  $\mathbf{h}_i=\mathbf{h}^0_i+\bm{\xi}_i(t)$ is the effective magnetic field, where $\bm{\xi}_i(t)$ is a three-dimensional(3D) stochastic magnetic field representing the thermal noise, which can be approximated as white noise if assuming that the typical time scale of the bath is much shorter than that of the system.  We further assume the local bath around each spin is independent with each other thus the stochastic variables satisfy:
\begin{eqnarray}
\langle \xi_i^\alpha(t)\rangle_{\bm\xi}&=&0 \\
\langle \xi_i^\alpha(t)\xi_j^\beta(t')\rangle_{\bm\xi}&=&\mathcal{D}^2\delta_{\alpha\beta}\delta_{ij} \delta(t-t')
\end{eqnarray}
where $\alpha,\beta=x,y,z$,  $\mathcal{D}$ is the strength of the noise, and the average $\langle\rangle_{\bm\xi}$ is over all the noise trajectories.  If the bath is in thermal equilibrium with temperature T, the fluctuation-dissipation theorem (FDT) indicates that the strengths of the friction and noise satisfy
\begin{equation}
\mathcal{D}^2=2T \lambda.
\end{equation}

Numerically, we adopt Stratonovich's formula of the stochastic differential Eq.(\ref{eq:EOM}), and solve it using the standard Heun  method\cite{Ament2016} with the time step of $\Delta t=10^{-3}$, the convergence of which has been checked numerically (see the Supplmentary Material(SM)\cite{Supplementary}).  The system size in our simulation ranges from $L=8$ to $L=28$, which enable us to systematically analyze the finite-size effect.  The ensemble average over the noise trajectories can be performed by directly sampling over $\mathcal{N}$ sets of noise realizations. ($\mathcal{N}$ ranges from $4\times 10^3$ to $10^5$ depending on the simulated system size). In our simulation, we calculate the evolution of the average magnetization along x-direction $M(t)=\langle\frac 1{L^3}\sum_i s_i^x(t)\rangle_{\bm \xi}$ to characterize various dynamical behaviors.  We choose the initial state as the ground state of Hamiltonian.(\ref{eq:ham}) with $t=0$, since we start from a spatially uniform initial state, $M(t)$ is proportional to the auto-correlation function $C(t)=\langle\frac 1{L^2}\sum_i s_i^x(0)s_i^x(t)\rangle_{\bm \xi}$, which characterizes the memory effect of the initial state information.  However, we find that the long-time behavior doesn't change even if we start from a non-uniform random initial state, that different spins will finally  synchronize with each other\cite{Supplementary}.

{\it Zero temperature phase diagram:  --} It is shown that even at zero temperature ($\mathcal{D}=0$),  the classical EOM.(\ref{eq:EOM}) could exhibit rich long-time dynamical behaviors, which can be found in the $V_0-\omega_0$ phase diagram as shown in Fig.1. For sufficiently large $V_0$ and small $\omega_0$, one can find a $DTC$-2 phase, where the ferromagnetic order parameter $M(t)$ oscillate with a period twice of that of the  driving ($\frac{2\pi}{\omega_0}$). Besides that, one can also find other DTC phases whose periods are other integer multiples of $\frac{2\pi}{\omega_0}$ ($DTC$-3 and  $DTC$-4 phases for instance). If both $V_0$ and $\omega_0$ are large enough, one can find a synchronization between $M(t)$ ad $V(t)$, both of which oscillates in the same period(synchronized phase).  Besides these periodic oscillations, one can also find other dynamical phases that $M(t)$ could decay to zero (paramagnetic phase), or oscillate chaotically (chaos).

\begin{figure*}[htb]
\includegraphics[width=0.325\linewidth,bb=95 55 737 518]{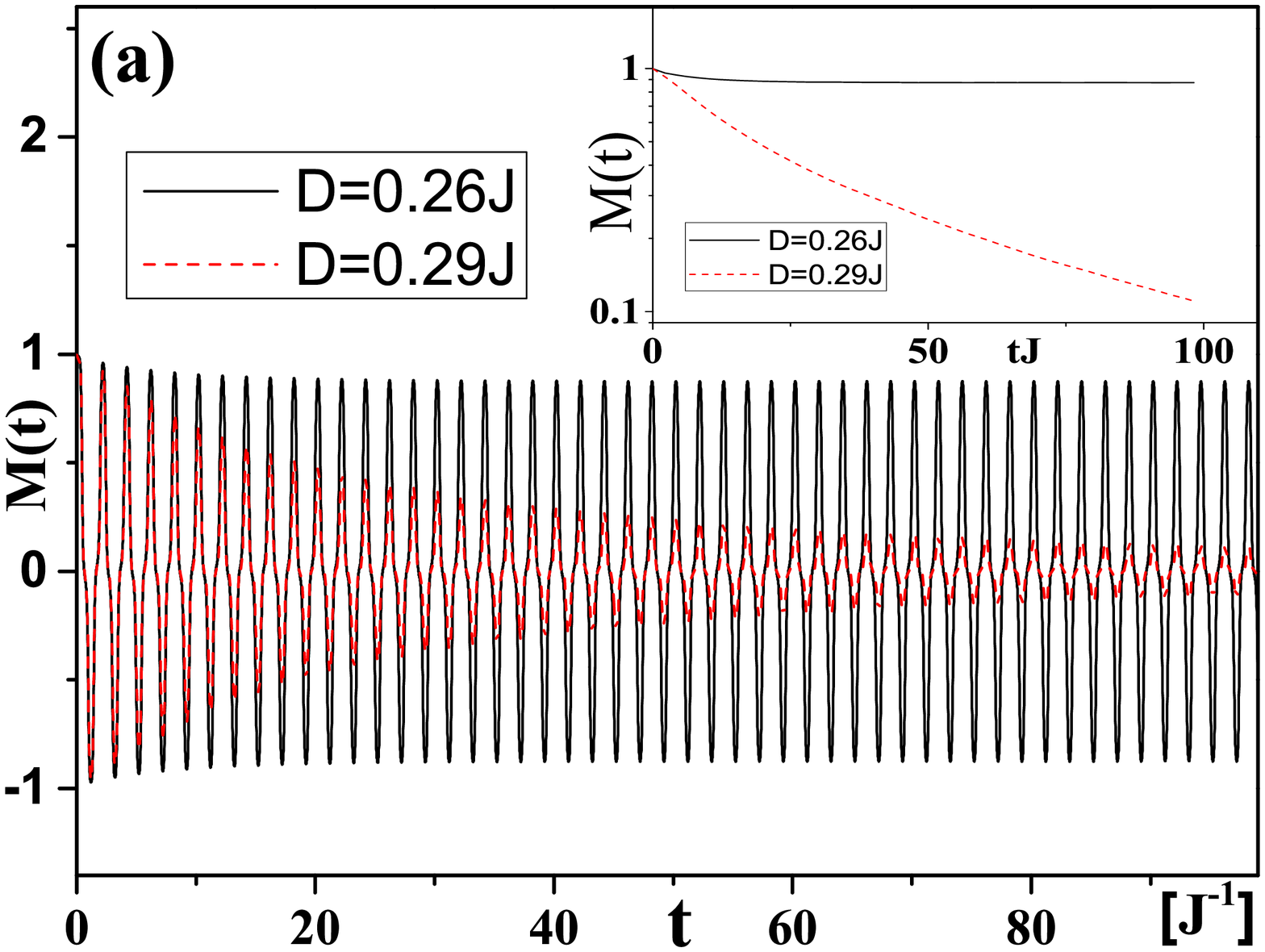}
\includegraphics[width=0.325\linewidth,bb=95 55 737 518]{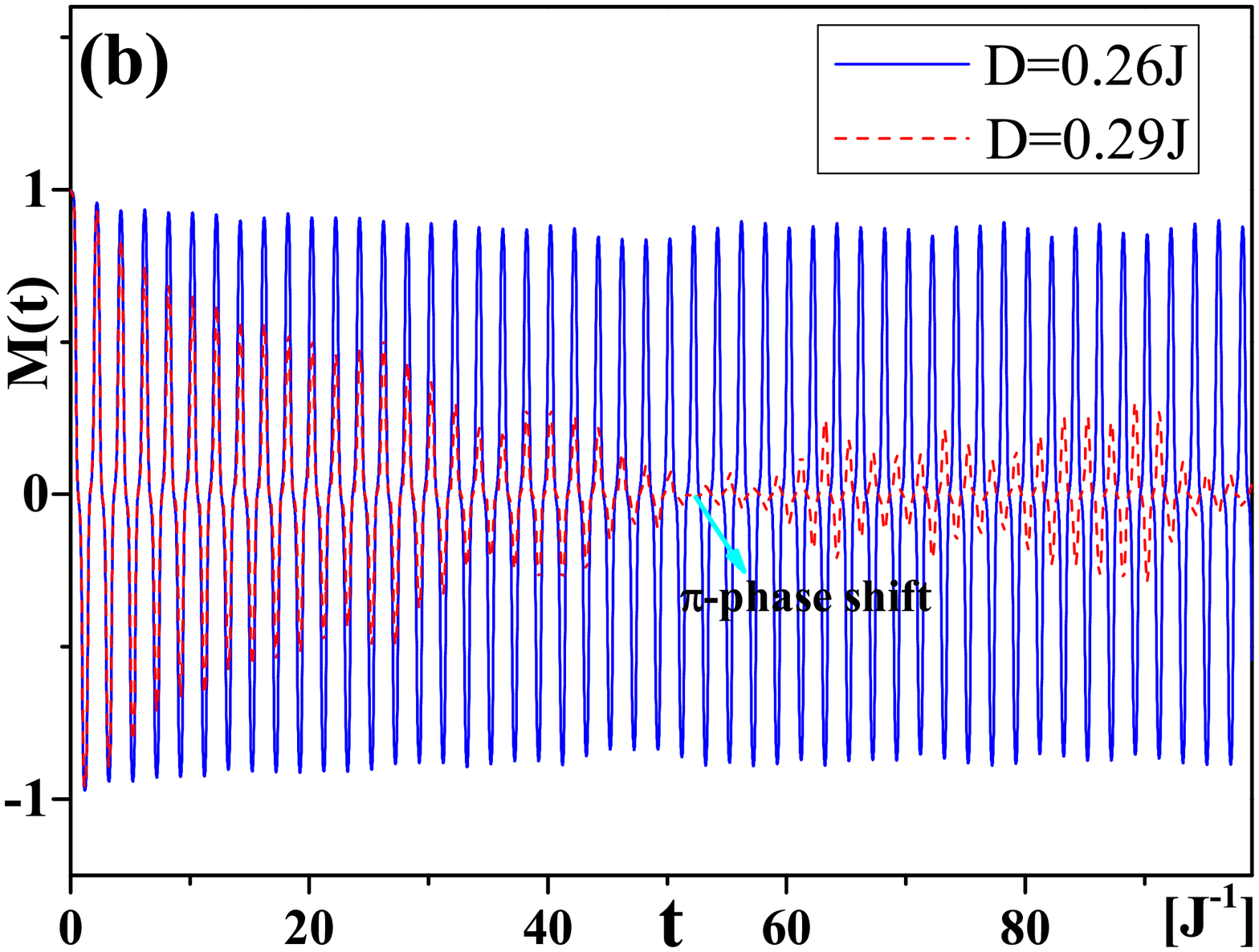}
\includegraphics[width=0.325\linewidth,bb=95 55 737 518]{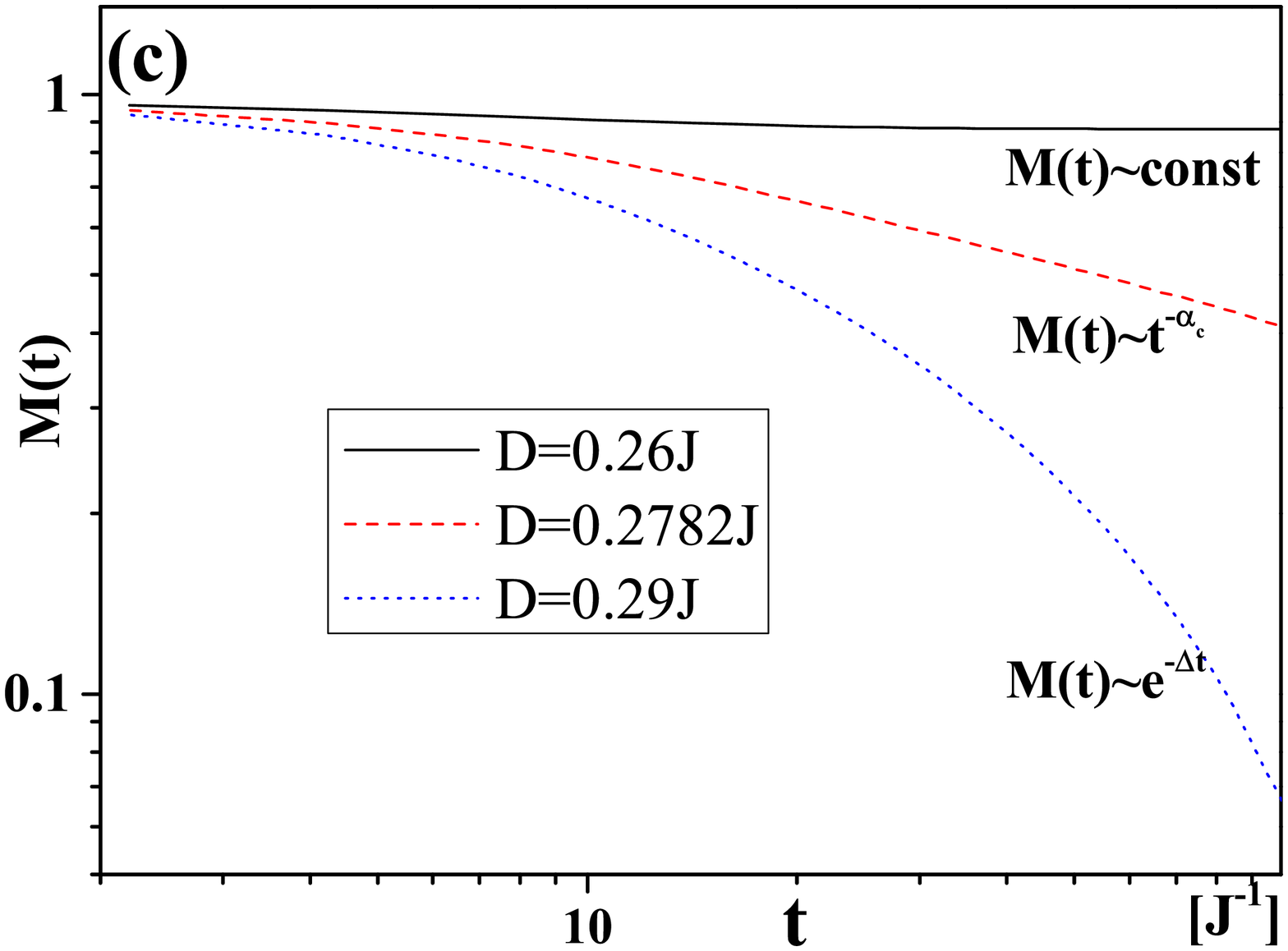}
\includegraphics[width=0.32\linewidth,bb=95 55 737 518]{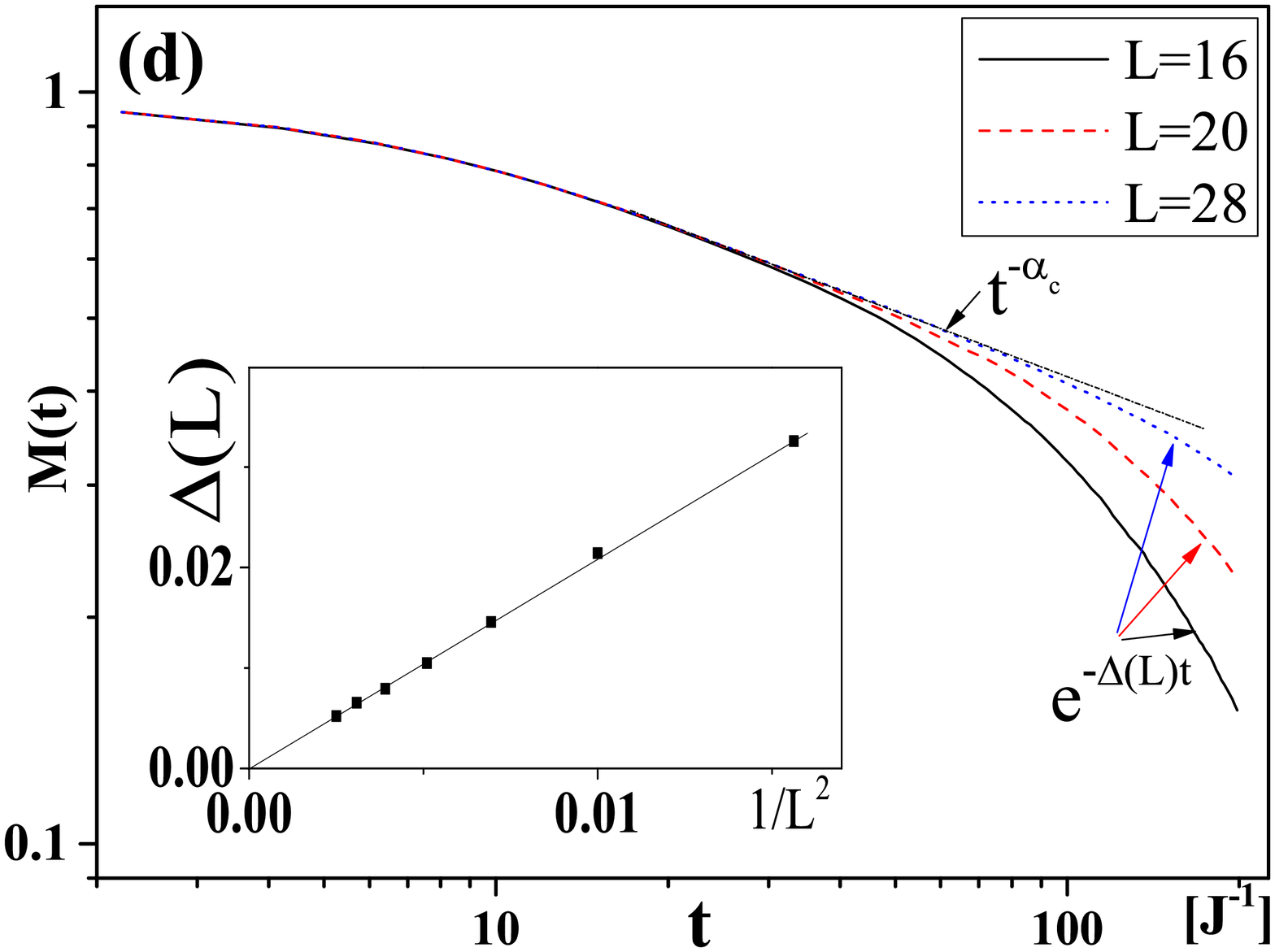}
\includegraphics[width=0.32\linewidth,bb=95 55 737 543]{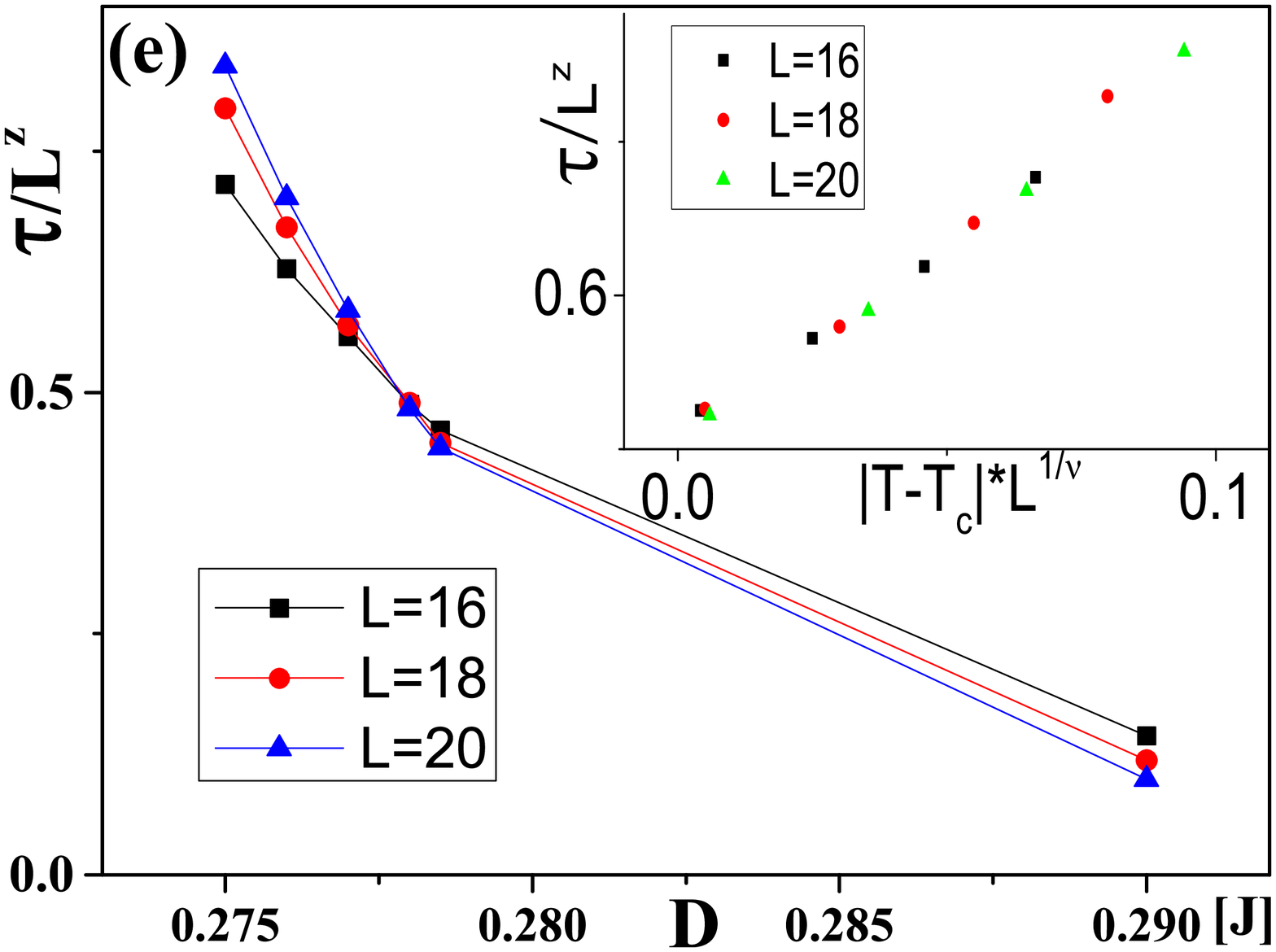}
\includegraphics[width=0.32\linewidth,bb=95 55 737 543]{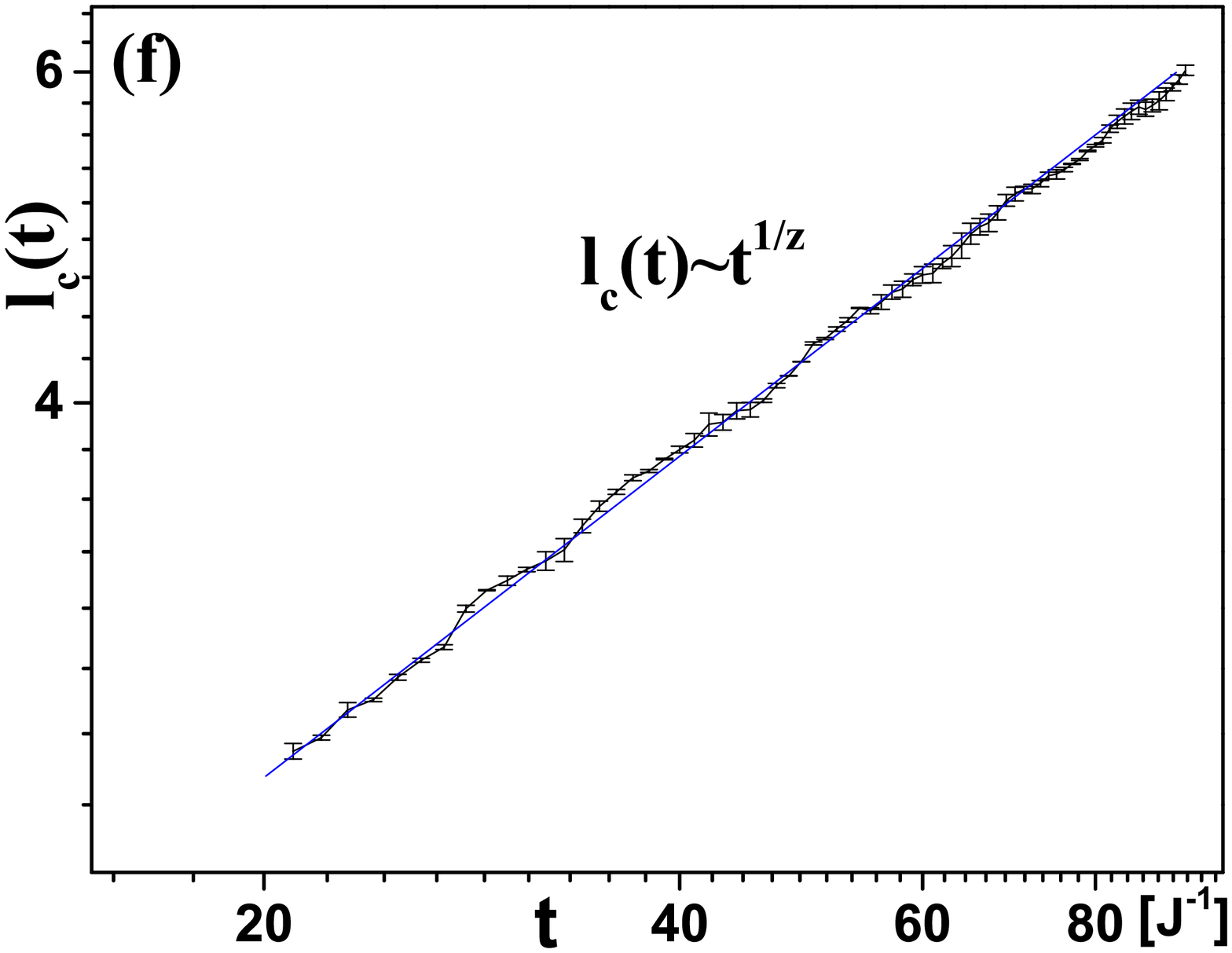}
\caption{(Color online). (a) The dynamics of the ferromagnetic order parameter $M(t)$ at temperatures above ($\mathcal{D}=0.29J$) and below ($\mathcal{D}=0.26J$) the critical temperature $T_c=\frac{D_c^2}{2J}$ with system size $L=20$. The inset is  the envelope (the peaks) of the oscillations in Fig.\ref{fig:fig2} (a) in a semi-log plot. (b) Comparison of $M(t)$ in the evolutions under a single noise trajectory with $\mathcal{D}=0.26J$ and $\mathcal{D}=0.29J$. A thermal activated $\pi$-phase shift can be found in the latter.  (c) Qualitatively different long-time behaviors of $M(t)$ at temperatures above, below and at the critical temperature.  (d)The dynamics of $M(t)$ at the critical point for systems with various system size $L$. The inset is the finite-size scaling of the exponents $\Delta(L)$ of late time exponential dynamics at $D=D_c$.  (e) The renormalized relaxation time as a function of $\mathcal{D}$ for systems with different system size $L$.  The insets are the data collapse for (e) using the critical exponents $\nu=0.64$ and $z=2$.  (f) At the critical temperature, the correlation length grows in time as $l_c(t)\sim t^{\frac 1z}$.  We fixed $V_0=5J$, $\omega_0=2\pi J$ and $h=\lambda=J$.  $L=20$ for (a) and (b), and $L=28$ for (c) and (f). $D=D_c=0.2782J$ for (d)-(f).
} \label{fig:fig2}
\end{figure*}

{\it The fate of DTC phase at various temperatures:  --} In the following, we will focus on the $DTC$-2 phase by fixing $V_0=5J$, $\omega_0=2\pi J$ and change the temperature. The noises, in general, will suppress the long-range order. As shown in Fig.\ref{fig:fig2} (a), $M(t)$ exhibits a damped oscillation whose amplitude decays exponentially with time ($\sim e^{-\Delta t}$ as shown in the inset) at a relatively high temperature (e.g $\mathcal{D}=0.29J$ corresponding to $T=0.042J$) . Such an exponential decay can be understood as a consequence of the ``kinks'' activated by thermal fluctuations:  the noise can induce tunnelings from one of the degenerate DTC phases to the other by a $\pi$-phase shifting\cite{Yang2021}, as shown in Fig.\ref{fig:fig2} (b) for the time evolution of M(t) under a single noise trajectory.  Such topological ``excitations'', no matter how rare they are, will inevitably destroy the long-range crystalline  order along the temporal direction via a similar mechanism responsible for the absence of spatial long-range order in 1D thermal equilibrium systems. At low temperature (e.g $\mathcal{D}=0.26J$ corresponding to $T=0.038J$), the thermal fluctuation is not strong enough to induce a $\pi$-phase shifting, but can only modulate the amplitude of the oscillation, thus we find a persistent oscillation of $M(t)$ as shown in Fig.\ref{fig:fig2} (a), which indicates a DTC phase with true long-range order in time domain.

{\it The critical behavior of the non-equilibrium phase transition--} As the temperature is lowered,  the exponent  of the exponential decay $\Delta$ decreases and finally vanishes when the temperature drops below a critical temperature $T_c=\frac{\mathcal{D}_c^2}{2J}$ with $D_c=0.2782J$, which indicates an absence of the exponential decay in the thermodynamic limit. For a sufficiently large system,  it is shown in Fig.\ref{fig:fig2} (c) that at the critical temperature ($T=T_c$), the amplitude of $M(t)$ decays algebraically with time ($\sim t^{-\alpha_c}$ with $\alpha_c=0.33(1)$),  qualitatively different from either the exponential decay at the high temperature or the persistent oscillation at low temperature. Such a critical exponent $\alpha_c=0.33$ is different from the value observed recently  ($\alpha_c=0.5$) in a mean-field analysis of the critical properties of a dissipative DTC phase\cite{Chinzei2021}.

For a finite system, the dynamics of $M(t)$ at the critical temperature can be separated into three regimes: a transient regime that depends on the initial state, a long-time exponential regime dominated by the finite size effect, and an intermediate universal algebraic regime sandwiched between them,  as shown in Fig.\ref{fig:fig2} (d).  The finite size effect leads to an exponential decay after sufficiently long time. However, as the system size increases, a finite size scaling of $\Delta_c(L)$  (the inset of Fig.\ref{fig:fig2} (d)) suggests that $\Delta_c(L)$ vanishes in the thermodynamic limit as $\Delta_c(L)\sim L^{-z}$ with $z=2$.  This result indicates a divergence of the relaxation time with the system size as $\tau_c\sim L^z$, where z is the dynamical critical exponent characterizing the dynamical universality class\cite{Hohenberg1977}. It is known that in a relaxation dynamics of a  critical kinetic Ising model without periodical driving, $z$ could be either $2$\cite{Glauber1963} or $3$\cite{Huse1986} depending on the conservation law during the relaxation dynamics.  Here, we find the dynamical critical exponent $z$ of our model  agrees well with the value in relaxation dynamics of undriven kinetic 2D Ising model without conservation law (Glauber model\cite{Glauber1963}). In the thermodynamic limit, $\Delta_c(L)$ vanishes and the exponential regime gives ways to the algebraic regime, which persists forever.

To study other critical exponents of this dynamical phase transition, we focus on the divergence of the relaxation time $\tau$ (the inverse of $\Delta$) close to the critical point. We first define a  renormalized relaxation time $\tilde{\tau}=\tau/L^z$ with $z=2$, and plot its dependence on $\mathcal{D}$ for systems with various system sizes $L$ around the critical point.  As shown in Fig.\ref{fig:fig2} (e),  one can find a scaling invariant point at $\mathcal{D}=\mathcal{D}_c$,  a signature of a continuous phase transition.  Near the critical point, the relaxation time diverges as $\tau\sim |T-T_c|^{-z\nu}$. To obtain the critical exponent $\nu$,   we perform the data collapses as shown in the insets of Fig.\ref{fig:fig2} (e),  from which we can extract that the critical exponents $\nu=0.64$, which agrees with the value ($\nu=0.642$) in the 3D Ising universality class .

Another important quantity to characterize the critical behavior is the correlation length $l$, which can be extracted from the equal-time  correlation function $S(\mathbf{r},t)=\langle s_1^x(t)s_\mathbf{r}^x(t)\rangle_{\bm \xi}- \langle s_1^x(t)\rangle_{\bm \xi} \langle s_\mathbf{r}^x(t)\rangle_{\bm \xi}$ as\cite{Sandvik2010}:
\begin{equation}
l_x(t)=\frac 1{|\mathbf{q}_x|}\sqrt{\frac{S(\mathbf{0},t)}{S(\mathbf{q}_x,t)}-1}
\end{equation}
where  $S(\mathbf{Q},t)=\frac{1}{L^3} \sum_{\mathbf{r}} e^{i\mathbf{Q}\cdot\mathbf{r}}S(\mathbf{r},t)$ is the structure factor of the spin-spin correlation. $\mathbf{q}_x=[\frac{2\pi}L,0,0]$ and $l_x(t)$ is the correlation length along x-direction at time t.  As shown in Fig.\ref{fig:fig2} (f), at the critical point, the correlation length diverges algebraically in time as $l(t)\sim t^{\beta}$ with $\beta= 0.55(3)$, which agrees with the dynamical critical exponent $\beta=1/z$ in the Glauber model.

{\it Discussion  --} Despite the presence of the periodical driving of our model, we find that its dynamical critical behaviors are similar with the relaxation dynamics of the undriven kinetic Ising model near the critical point (the model A in Hohenberg-Halperin's alphabetic classification of dynamical critical phenomena\cite{Hohenberg1977}). This result can be understood by a time-independent description of stroboscopic dynamics under periodic driving. Such a time-independent Hamiltonian is similar with the Floquet Hamiltonian in the periodically-driven quantum systems, and can be derived using an  approximation scheme similar with the Magnus expansion, a series expansion in terms of the driving period\cite{Magnus1954}. Even though  higher order terms (the four-site or multi-site interactions for instance)  may appear in the effective time-independent Hamiltonian, they could be irrelevant for the long-time dynamics near the critical point, thus the stroboscopic dynamics of our periodically driven system is similar with the relaxation dynamics of a time-independent kinetic Ising model near the critical point.

This similarity reminds us of the dynamical phase transition observed in a kinetic Ising model in a periodically oscillating magnetic field\cite{Korniss2000}, which also experiences a non-equilibrium phase transition from a symmetry-restoring oscillations to  symmetry-breaking oscillations for the time-dependent magnetization, where the dynamical critical behavior is also found to be the same as those of the undriven cases\cite{Fujisaka2001}.  In spite of the similarity of this model\cite{Korniss2000} and our model,  there is an important difference that in our model, instead of external field, the periodic driving is imposed on the strength of interaction, which give rise to the consequence of the spontaneous breaking of discrete TTS.  While in model with oscillating field\cite{Korniss2000}, the period of the magnetization and external field is always the same, thus the discrete TTS is not broken, corresponding to ``synchronized phase'' in our language.

Another important issue has not been discussed so far is the quantum effect: what will happen if thermal noises are replaced by quantum fluctuations? In another word, if the Hamiltonian.(\ref{eq:ham}) is a quantum transverse Ising model, can we observe DTC behavior in such a periodical-driven quantum system? The effect of quantum fluctuations on systems with continuous TTS breaking has been studied in the single-particle level.  For instance, for a Van der Pol oscillator with a classical trajectory of limiting circle,  quantum fluctuation induces a phase diffusion in the limiting circle\cite{Navarrete2017}, which recovers the continuous TTS symmetry thus destroys the long-range TC order. On the contrary, our model breaks the discrete TTS, which is expected to be more stable compared to the continuous ones. The competition between the quantum fluctuations and time crystalline order may give rise to rich dynamical phenomena, thus is worthwhile to be explored in the future work, even though a numerical simulation of such a non-equilibrium 3D (open) quantum many-body system is a formidable, if not impossible, task.

{\it Conclusion and outlook  --} In conclusion, we study the stability of  DTC phase in the presence of thermal bath, and find that despite its genuine non-equilibrium feature, this DTC phase shares a lot of common properties with equilibrium ordered phases with spatially $Z_2$ symmetry breaking, especially its finite-size effect and critical behavior.   Future developments will include an analytic explanation of the critical behavior, which calls for a coarse-grained effective description and a non-equilibrium field theory analysis of our system in the Keldysh formalism\cite{Martin1973,Kamenev2011,Altland2010}. Recently, Natsheh {\it et al} analytically studied the critical behavior of a periodically driven quantum O(N) model with DTC phase\cite{Natsheh2021}, a generalization of this work may shed light on our problem. Another important question involves  the generality of our conclusion: whether similar behavior can be observed in other DTC models (e.g. the coupled driven-dissipative nonlinear pendulums\cite{Yao2020})? Last but not the least, whether it is possible to find the melting transitions with the dynamical universality that go beyond the undriven cases? After all, the Magnus expansion could generate long-range interactions in the effective Hamiltonian, especially in the presence of low-frequency driving, and it is possible that such long-range interaction may qualitatively change the critical behavior of the systems and give rise to intriguing dynamical universality classes.

{\it Acknowledgments}.---This work is supported by the National Key Research and Development Program of China (Grant No.2020YFA0309000), Natural Science Foundation of  China (Grant No.12174251),  Natural Science Foundation of Shanghai (Grant No.22ZR142830),   Shanghai Municipal Science and Technology Major Project (Grant No.2019SHZDZX01).


\begin{thebibliography}{57}
\expandafter\ifx\csname natexlab\endcsname\relax\def\natexlab#1{#1}\fi
\expandafter\ifx\csname bibnamefont\endcsname\relax
  \def\bibnamefont#1{#1}\fi
\expandafter\ifx\csname bibfnamefont\endcsname\relax
  \def\bibfnamefont#1{#1}\fi
\expandafter\ifx\csname citenamefont\endcsname\relax
  \def\citenamefont#1{#1}\fi
\expandafter\ifx\csname url\endcsname\relax
  \def\url#1{\texttt{#1}}\fi
\expandafter\ifx\csname urlprefix\endcsname\relax\def\urlprefix{URL }\fi
\providecommand{\bibinfo}[2]{#2}
\providecommand{\eprint}[2][]{\url{#2}}

\bibitem[{\citenamefont{Wilczek}(2012)}]{Wilczek2012}
\bibinfo{author}{\bibfnamefont{F.}~\bibnamefont{Wilczek}},
  \bibinfo{journal}{Phys. Rev. Lett.} \textbf{\bibinfo{volume}{109}},
  \bibinfo{pages}{160401} (\bibinfo{year}{2012}).

\bibitem[{\citenamefont{Hohenberg and Halperin}(1977)}]{Hohenberg1977}
\bibinfo{author}{\bibfnamefont{P.~C.} \bibnamefont{Hohenberg}}
  \bibnamefont{and} \bibinfo{author}{\bibfnamefont{B.~I.}
  \bibnamefont{Halperin}}, \bibinfo{journal}{Rev. Mod. Phys.}
  \textbf{\bibinfo{volume}{49}}, \bibinfo{pages}{435} (\bibinfo{year}{1977}).

\bibitem[{\citenamefont{Taeuber}(2014)}]{Taeuber2014}
\bibinfo{author}{\bibfnamefont{U.}~\bibnamefont{Taeuber}},
  \emph{\bibinfo{title}{CRITICAL DYNAMICS: A Field Theory Approach to
  Equilibrium and Non-Equilibrium Scaling Behavior}}
  (\bibinfo{publisher}{~Cambridge University Press, Cambridge},
  \bibinfo{year}{2014}).

\bibitem[{\citenamefont{Kardar et~al.}(1986)\citenamefont{Kardar, Parisi, and
  Zhang}}]{Kardar1986}
\bibinfo{author}{\bibfnamefont{M.}~\bibnamefont{Kardar}},
  \bibinfo{author}{\bibfnamefont{G.}~\bibnamefont{Parisi}}, \bibnamefont{and}
  \bibinfo{author}{\bibfnamefont{Y.-C.} \bibnamefont{Zhang}},
  \bibinfo{journal}{Phys. Rev. Lett.} \textbf{\bibinfo{volume}{56}},
  \bibinfo{pages}{889} (\bibinfo{year}{1986}).

\bibitem[{\citenamefont{Shapere and Wilczek}(2012)}]{Shapere2012}
\bibinfo{author}{\bibfnamefont{A.}~\bibnamefont{Shapere}} \bibnamefont{and}
  \bibinfo{author}{\bibfnamefont{F.}~\bibnamefont{Wilczek}},
  \bibinfo{journal}{Phys. Rev. Lett.} \textbf{\bibinfo{volume}{109}},
  \bibinfo{pages}{160402} (\bibinfo{year}{2012}).

\bibitem[{\citenamefont{Li et~al.}(2012)\citenamefont{Li, Gong, Yin, Quan, Yin,
  Zhang, Duan, and Zhang}}]{Li2012}
\bibinfo{author}{\bibfnamefont{T.}~\bibnamefont{Li}},
  \bibinfo{author}{\bibfnamefont{Z.-X.} \bibnamefont{Gong}},
  \bibinfo{author}{\bibfnamefont{Z.-Q.} \bibnamefont{Yin}},
  \bibinfo{author}{\bibfnamefont{H.~T.} \bibnamefont{Quan}},
  \bibinfo{author}{\bibfnamefont{X.}~\bibnamefont{Yin}},
  \bibinfo{author}{\bibfnamefont{P.}~\bibnamefont{Zhang}},
  \bibinfo{author}{\bibfnamefont{L.-M.} \bibnamefont{Duan}}, \bibnamefont{and}
  \bibinfo{author}{\bibfnamefont{X.}~\bibnamefont{Zhang}},
  \bibinfo{journal}{Phys. Rev. Lett.} \textbf{\bibinfo{volume}{109}},
  \bibinfo{pages}{163001} (\bibinfo{year}{2012}).

\bibitem[{\citenamefont{Wilczek}(2013)}]{Wilczek2013}
\bibinfo{author}{\bibfnamefont{F.}~\bibnamefont{Wilczek}},
  \bibinfo{journal}{Phys. Rev. Lett.} \textbf{\bibinfo{volume}{111}},
  \bibinfo{pages}{250402} (\bibinfo{year}{2013}).

\bibitem[{\citenamefont{Sacha}(2015)}]{Sacha2015}
\bibinfo{author}{\bibfnamefont{K.}~\bibnamefont{Sacha}},
  \bibinfo{journal}{Phys. Rev. A} \textbf{\bibinfo{volume}{91}},
  \bibinfo{pages}{033617} (\bibinfo{year}{2015}).

\bibitem[{\citenamefont{Else et~al.}(2016)\citenamefont{Else, Bauer, and
  Nayak}}]{Else2016}
\bibinfo{author}{\bibfnamefont{D.~V.} \bibnamefont{Else}},
  \bibinfo{author}{\bibfnamefont{B.}~\bibnamefont{Bauer}}, \bibnamefont{and}
  \bibinfo{author}{\bibfnamefont{C.}~\bibnamefont{Nayak}},
  \bibinfo{journal}{Phys. Rev. Lett.} \textbf{\bibinfo{volume}{117}},
  \bibinfo{pages}{090402} (\bibinfo{year}{2016}).

\bibitem[{\citenamefont{Khemani et~al.}(2016)\citenamefont{Khemani, Lazarides,
  Moessner, and Sondhi}}]{Khemani2016}
\bibinfo{author}{\bibfnamefont{V.}~\bibnamefont{Khemani}},
  \bibinfo{author}{\bibfnamefont{A.}~\bibnamefont{Lazarides}},
  \bibinfo{author}{\bibfnamefont{R.}~\bibnamefont{Moessner}}, \bibnamefont{and}
  \bibinfo{author}{\bibfnamefont{S.~L.} \bibnamefont{Sondhi}},
  \bibinfo{journal}{Phys. Rev. Lett.} \textbf{\bibinfo{volume}{116}},
  \bibinfo{pages}{250401} (\bibinfo{year}{2016}).

\bibitem[{\citenamefont{Yao et~al.}(2017)\citenamefont{Yao, Potter, Potirniche,
  and Vishwanath}}]{Yao2017}
\bibinfo{author}{\bibfnamefont{N.~Y.} \bibnamefont{Yao}},
  \bibinfo{author}{\bibfnamefont{A.~C.} \bibnamefont{Potter}},
  \bibinfo{author}{\bibfnamefont{I.-D.} \bibnamefont{Potirniche}},
  \bibnamefont{and}
  \bibinfo{author}{\bibfnamefont{A.}~\bibnamefont{Vishwanath}},
  \bibinfo{journal}{Phys. Rev. Lett.} \textbf{\bibinfo{volume}{118}},
  \bibinfo{pages}{030401} (\bibinfo{year}{2017}).

\bibitem[{\citenamefont{Russomanno et~al.}(2017)\citenamefont{Russomanno,
  Iemini, Dalmonte, and Fazio}}]{Russomanno2017}
\bibinfo{author}{\bibfnamefont{A.}~\bibnamefont{Russomanno}},
  \bibinfo{author}{\bibfnamefont{F.}~\bibnamefont{Iemini}},
  \bibinfo{author}{\bibfnamefont{M.}~\bibnamefont{Dalmonte}}, \bibnamefont{and}
  \bibinfo{author}{\bibfnamefont{R.}~\bibnamefont{Fazio}},
  \bibinfo{journal}{Phys. Rev. B} \textbf{\bibinfo{volume}{95}},
  \bibinfo{pages}{214307} (\bibinfo{year}{2017}).

\bibitem[{\citenamefont{Gong et~al.}(2018)\citenamefont{Gong, Hamazaki, and
  Ueda}}]{Gong2018}
\bibinfo{author}{\bibfnamefont{Z.}~\bibnamefont{Gong}},
  \bibinfo{author}{\bibfnamefont{R.}~\bibnamefont{Hamazaki}}, \bibnamefont{and}
  \bibinfo{author}{\bibfnamefont{M.}~\bibnamefont{Ueda}},
  \bibinfo{journal}{Phys. Rev. Lett.} \textbf{\bibinfo{volume}{120}},
  \bibinfo{pages}{040404} (\bibinfo{year}{2018}).

\bibitem[{\citenamefont{Huang et~al.}(2018)\citenamefont{Huang, Wu, and
  Liu}}]{Huang2018}
\bibinfo{author}{\bibfnamefont{B.}~\bibnamefont{Huang}},
  \bibinfo{author}{\bibfnamefont{Y.-H.} \bibnamefont{Wu}}, \bibnamefont{and}
  \bibinfo{author}{\bibfnamefont{W.~V.} \bibnamefont{Liu}},
  \bibinfo{journal}{Phys. Rev. Lett.} \textbf{\bibinfo{volume}{120}},
  \bibinfo{pages}{110603} (\bibinfo{year}{2018}).

\bibitem[{\citenamefont{Iemini et~al.}(2018)\citenamefont{Iemini, Russomanno,
  Keeling, Schir\`o, Dalmonte, and Fazio}}]{Iemini2018}
\bibinfo{author}{\bibfnamefont{F.}~\bibnamefont{Iemini}},
  \bibinfo{author}{\bibfnamefont{A.}~\bibnamefont{Russomanno}},
  \bibinfo{author}{\bibfnamefont{J.}~\bibnamefont{Keeling}},
  \bibinfo{author}{\bibfnamefont{M.}~\bibnamefont{Schir\`o}},
  \bibinfo{author}{\bibfnamefont{M.}~\bibnamefont{Dalmonte}}, \bibnamefont{and}
  \bibinfo{author}{\bibfnamefont{R.}~\bibnamefont{Fazio}},
  \bibinfo{journal}{Phys. Rev. Lett.} \textbf{\bibinfo{volume}{121}},
  \bibinfo{pages}{035301} (\bibinfo{year}{2018}).

\bibitem[{\citenamefont{Das et~al.}(2018)\citenamefont{Das, Pan, Ghosh, and
  Pal}}]{Das2018}
\bibinfo{author}{\bibfnamefont{P.}~\bibnamefont{Das}},
  \bibinfo{author}{\bibfnamefont{S.}~\bibnamefont{Pan}},
  \bibinfo{author}{\bibfnamefont{S.}~\bibnamefont{Ghosh}}, \bibnamefont{and}
  \bibinfo{author}{\bibfnamefont{P.}~\bibnamefont{Pal}},
  \bibinfo{journal}{Phys. Rev. D} \textbf{\bibinfo{volume}{98}},
  \bibinfo{pages}{024004} (\bibinfo{year}{2018}).

\bibitem[{\citenamefont{Zhu et~al.}(2019)\citenamefont{Zhu, Marino, Yao, Lukin,
  and Demler}}]{Zhu2019}
\bibinfo{author}{\bibfnamefont{B.}~\bibnamefont{Zhu}},
  \bibinfo{author}{\bibfnamefont{J.}~\bibnamefont{Marino}},
  \bibinfo{author}{\bibfnamefont{N.~Y.} \bibnamefont{Yao}},
  \bibinfo{author}{\bibfnamefont{M.~D.} \bibnamefont{Lukin}}, \bibnamefont{and}
  \bibinfo{author}{\bibfnamefont{E.~A.} \bibnamefont{Demler}},
  \bibinfo{journal}{New Journal of Physics} \textbf{\bibinfo{volume}{21}},
  \bibinfo{pages}{073028} (\bibinfo{year}{2019}).

\bibitem[{\citenamefont{Liao et~al.}(2019)\citenamefont{Liao, Smits, van~der
  Straten, and Stoof}}]{Liao2019}
\bibinfo{author}{\bibfnamefont{L.}~\bibnamefont{Liao}},
  \bibinfo{author}{\bibfnamefont{J.}~\bibnamefont{Smits}},
  \bibinfo{author}{\bibfnamefont{P.}~\bibnamefont{van~der Straten}},
  \bibnamefont{and} \bibinfo{author}{\bibfnamefont{H.~T.~C.}
  \bibnamefont{Stoof}}, \bibinfo{journal}{Phys. Rev. A}
  \textbf{\bibinfo{volume}{99}}, \bibinfo{pages}{013625}
  (\bibinfo{year}{2019}).

\bibitem[{\citenamefont{Kozin and Kyriienko}(2019)}]{Kozin2019}
\bibinfo{author}{\bibfnamefont{V.~K.} \bibnamefont{Kozin}} \bibnamefont{and}
  \bibinfo{author}{\bibfnamefont{O.}~\bibnamefont{Kyriienko}},
  \bibinfo{journal}{Phys. Rev. Lett.} \textbf{\bibinfo{volume}{123}},
  \bibinfo{pages}{210602} (\bibinfo{year}{2019}).

\bibitem[{\citenamefont{Cai et~al.}(2020)\citenamefont{Cai, Huang, and
  Liu}}]{Cai2020}
\bibinfo{author}{\bibfnamefont{Z.}~\bibnamefont{Cai}},
  \bibinfo{author}{\bibfnamefont{Y.}~\bibnamefont{Huang}}, \bibnamefont{and}
  \bibinfo{author}{\bibfnamefont{W.~V.} \bibnamefont{Liu}},
  \bibinfo{journal}{Chinese Physics Letters} \textbf{\bibinfo{volume}{37}},
  \bibinfo{pages}{050503} (\bibinfo{year}{2020}).

\bibitem[{\citenamefont{Chinzei and Ikeda}(2020)}]{Chinzei2020}
\bibinfo{author}{\bibfnamefont{K.}~\bibnamefont{Chinzei}} \bibnamefont{and}
  \bibinfo{author}{\bibfnamefont{T.~N.} \bibnamefont{Ikeda}},
  \bibinfo{journal}{Phys. Rev. Lett.} \textbf{\bibinfo{volume}{125}},
  \bibinfo{pages}{060601} (\bibinfo{year}{2020}).

\bibitem[{\citenamefont{Lyu et~al.}(2020)\citenamefont{Lyu, Choudhury, Lv, Yan,
  and Zhou}}]{Lyu2020}
\bibinfo{author}{\bibfnamefont{C.}~\bibnamefont{Lyu}},
  \bibinfo{author}{\bibfnamefont{S.}~\bibnamefont{Choudhury}},
  \bibinfo{author}{\bibfnamefont{C.}~\bibnamefont{Lv}},
  \bibinfo{author}{\bibfnamefont{Y.}~\bibnamefont{Yan}}, \bibnamefont{and}
  \bibinfo{author}{\bibfnamefont{Q.}~\bibnamefont{Zhou}},
  \bibinfo{journal}{Phys. Rev. Research} \textbf{\bibinfo{volume}{2}},
  \bibinfo{pages}{033070} (\bibinfo{year}{2020}).

\bibitem[{\citenamefont{{Choudhury} and {Liu}}(2021)}]{Choudhury2021}
\bibinfo{author}{\bibfnamefont{S.}~\bibnamefont{{Choudhury}}} \bibnamefont{and}
  \bibinfo{author}{\bibfnamefont{W.~V.} \bibnamefont{{Liu}}},
  \bibinfo{journal}{arXiv e-prints} \bibinfo{eid}{arXiv:2109.05318}
  (\bibinfo{year}{2021}), \eprint{2109.05318}.

\bibitem[{\citenamefont{Bruno}(2013)}]{Bruno2013}
\bibinfo{author}{\bibfnamefont{P.}~\bibnamefont{Bruno}},
  \bibinfo{journal}{Phys. Rev. Lett.} \textbf{\bibinfo{volume}{111}},
  \bibinfo{pages}{070402} (\bibinfo{year}{2013}).

\bibitem[{\citenamefont{Watanabe and Oshikawa}(2015)}]{Watanabe2015}
\bibinfo{author}{\bibfnamefont{H.}~\bibnamefont{Watanabe}} \bibnamefont{and}
  \bibinfo{author}{\bibfnamefont{M.}~\bibnamefont{Oshikawa}},
  \bibinfo{journal}{Phys. Rev. Lett.} \textbf{\bibinfo{volume}{114}},
  \bibinfo{pages}{251603} (\bibinfo{year}{2015}).

\bibitem[{\citenamefont{Choi et~al.}(2017)\citenamefont{Choi, Landig, Kucsko,
  Zhou, Isoya, Jelezko, Onoda, Sumiya, Khemani, von Keyserlingk
  et~al.}}]{Choi2017}
\bibinfo{author}{\bibfnamefont{S.}~\bibnamefont{Choi}},
  \bibinfo{author}{\bibfnamefont{R.}~\bibnamefont{Landig}},
  \bibinfo{author}{\bibfnamefont{G.}~\bibnamefont{Kucsko}},
  \bibinfo{author}{\bibfnamefont{H.}~\bibnamefont{Zhou}},
  \bibinfo{author}{\bibfnamefont{J.}~\bibnamefont{Isoya}},
  \bibinfo{author}{\bibfnamefont{F.}~\bibnamefont{Jelezko}},
  \bibinfo{author}{\bibfnamefont{S.}~\bibnamefont{Onoda}},
  \bibinfo{author}{\bibfnamefont{H.}~\bibnamefont{Sumiya}},
  \bibinfo{author}{\bibfnamefont{V.}~\bibnamefont{Khemani}},
  \bibinfo{author}{\bibfnamefont{C.}~\bibnamefont{von Keyserlingk}},
  \bibnamefont{et~al.}, \bibinfo{journal}{Nature}
  \textbf{\bibinfo{volume}{543}}, \bibinfo{pages}{221} (\bibinfo{year}{2017}).

\bibitem[{\citenamefont{Zhang et~al.}(2017)\citenamefont{Zhang, Hess,
  Kyprianidis, Becker, Lee, Smith, Pagano, Potirniche, Potter, Vishwanath
  et~al.}}]{Zhang2017}
\bibinfo{author}{\bibfnamefont{J.}~\bibnamefont{Zhang}},
  \bibinfo{author}{\bibfnamefont{P.~W.} \bibnamefont{Hess}},
  \bibinfo{author}{\bibfnamefont{A.}~\bibnamefont{Kyprianidis}},
  \bibinfo{author}{\bibfnamefont{P.}~\bibnamefont{Becker}},
  \bibinfo{author}{\bibfnamefont{A.}~\bibnamefont{Lee}},
  \bibinfo{author}{\bibfnamefont{J.}~\bibnamefont{Smith}},
  \bibinfo{author}{\bibfnamefont{G.}~\bibnamefont{Pagano}},
  \bibinfo{author}{\bibfnamefont{I.-D.} \bibnamefont{Potirniche}},
  \bibinfo{author}{\bibfnamefont{A.~C.} \bibnamefont{Potter}},
  \bibinfo{author}{\bibfnamefont{A.}~\bibnamefont{Vishwanath}},
  \bibnamefont{et~al.}, \bibinfo{journal}{Nature}
  \textbf{\bibinfo{volume}{543}}, \bibinfo{pages}{217} (\bibinfo{year}{2017}).

\bibitem[{\citenamefont{Autti et~al.}(2018)\citenamefont{Autti, Eltsov, and
  Volovik}}]{Autti2018}
\bibinfo{author}{\bibfnamefont{S.}~\bibnamefont{Autti}},
  \bibinfo{author}{\bibfnamefont{V.~B.} \bibnamefont{Eltsov}},
  \bibnamefont{and} \bibinfo{author}{\bibfnamefont{G.~E.}
  \bibnamefont{Volovik}}, \bibinfo{journal}{Phys. Rev. Lett.}
  \textbf{\bibinfo{volume}{120}}, \bibinfo{pages}{215301}
  (\bibinfo{year}{2018}).

\bibitem[{\citenamefont{Tr\"ager et~al.}(2021)\citenamefont{Tr\"ager,
  Gruszecki, Lisiecki, Gro\ss{}, F\"orster, Weigand,
  G\l{}owi\ifmmode~\acute{n}\else \'{n}\fi{}ski, Ku\ifmmode~\acute{s}\else
  \'{s}\fi{}wik, Dubowik, Sch\"utz et~al.}}]{Trager2021}
\bibinfo{author}{\bibfnamefont{N.}~\bibnamefont{Tr\"ager}},
  \bibinfo{author}{\bibfnamefont{P.}~\bibnamefont{Gruszecki}},
  \bibinfo{author}{\bibfnamefont{F.}~\bibnamefont{Lisiecki}},
  \bibinfo{author}{\bibfnamefont{F.}~\bibnamefont{Gro\ss{}}},
  \bibinfo{author}{\bibfnamefont{J.}~\bibnamefont{F\"orster}},
  \bibinfo{author}{\bibfnamefont{M.}~\bibnamefont{Weigand}},
  \bibinfo{author}{\bibfnamefont{H.}~\bibnamefont{G\l{}owi\ifmmode~\acute{n}\else
  \'{n}\fi{}ski}},
  \bibinfo{author}{\bibfnamefont{P.}~\bibnamefont{Ku\ifmmode~\acute{s}\else
  \'{s}\fi{}wik}}, \bibinfo{author}{\bibfnamefont{J.}~\bibnamefont{Dubowik}},
  \bibinfo{author}{\bibfnamefont{G.}~\bibnamefont{Sch\"utz}},
  \bibnamefont{et~al.}, \bibinfo{journal}{Phys. Rev. Lett.}
  \textbf{\bibinfo{volume}{126}}, \bibinfo{pages}{057201}
  (\bibinfo{year}{2021}).

\bibitem[{\citenamefont{Ke\ss{}ler et~al.}(2021)\citenamefont{Ke\ss{}ler,
  Kongkhambut, Georges, Mathey, Cosme, and Hemmerich}}]{Kessler2021}
\bibinfo{author}{\bibfnamefont{H.}~\bibnamefont{Ke\ss{}ler}},
  \bibinfo{author}{\bibfnamefont{P.}~\bibnamefont{Kongkhambut}},
  \bibinfo{author}{\bibfnamefont{C.}~\bibnamefont{Georges}},
  \bibinfo{author}{\bibfnamefont{L.}~\bibnamefont{Mathey}},
  \bibinfo{author}{\bibfnamefont{J.~G.} \bibnamefont{Cosme}}, \bibnamefont{and}
  \bibinfo{author}{\bibfnamefont{A.}~\bibnamefont{Hemmerich}},
  \bibinfo{journal}{Phys. Rev. Lett.} \textbf{\bibinfo{volume}{127}},
  \bibinfo{pages}{043602} (\bibinfo{year}{2021}).

\bibitem[{\citenamefont{{Mi} et~al.}(2021)\citenamefont{{Mi}, {Ippoliti},
  {Quintana}, {Greene}, {Chen}, {Gross}, {Arute}, {Arya}, {Atalaya}, {Babbush}
  et~al.}}]{Mi2021}
\bibinfo{author}{\bibfnamefont{X.}~\bibnamefont{{Mi}}},
  \bibinfo{author}{\bibfnamefont{M.}~\bibnamefont{{Ippoliti}}},
  \bibinfo{author}{\bibfnamefont{C.}~\bibnamefont{{Quintana}}},
  \bibinfo{author}{\bibfnamefont{A.}~\bibnamefont{{Greene}}},
  \bibinfo{author}{\bibfnamefont{Z.}~\bibnamefont{{Chen}}},
  \bibinfo{author}{\bibfnamefont{J.}~\bibnamefont{{Gross}}},
  \bibinfo{author}{\bibfnamefont{F.}~\bibnamefont{{Arute}}},
  \bibinfo{author}{\bibfnamefont{K.}~\bibnamefont{{Arya}}},
  \bibinfo{author}{\bibfnamefont{J.}~\bibnamefont{{Atalaya}}},
  \bibinfo{author}{\bibfnamefont{R.}~\bibnamefont{{Babbush}}},
  \bibnamefont{et~al.}, \bibinfo{journal}{Nature}
  \textbf{\bibinfo{volume}{601}}, \bibinfo{pages}{531} (\bibinfo{year}{2022}).


\bibitem[{\citenamefont{{Frey} and {Rachel}}(2021)}]{Frey2021}
\bibinfo{author}{\bibfnamefont{P.}~\bibnamefont{{Frey}}} \bibnamefont{and}
  \bibinfo{author}{\bibfnamefont{S.}~\bibnamefont{{Rachel}}},
  \bibinfo{journal}{arXiv e-prints} \bibinfo{eid}{arXiv:2105.06632}
  (\bibinfo{year}{2021}), \eprint{2105.06632}.

\bibitem[{\citenamefont{Yao et~al.}(2020)\citenamefont{Yao, Nayak, Balents, and
  Zaletel}}]{Yao2020}
\bibinfo{author}{\bibfnamefont{N.~Y.} \bibnamefont{Yao}},
  \bibinfo{author}{\bibfnamefont{C.}~\bibnamefont{Nayak}},
  \bibinfo{author}{\bibfnamefont{L.}~\bibnamefont{Balents}}, \bibnamefont{and}
  \bibinfo{author}{\bibfnamefont{M.~P.} \bibnamefont{Zaletel}},
  \bibinfo{journal}{Nat. Phys.} \textbf{\bibinfo{volume}{16}},
  \bibinfo{pages}{438} (\bibinfo{year}{2020}).

\bibitem[{\citenamefont{{Zhuang} et~al.}(2021)\citenamefont{{Zhuang},
  {Machado}, {Yao}, and {Zaletel}}}]{Zhuang2021}
\bibinfo{author}{\bibfnamefont{Q.}~\bibnamefont{{Zhuang}}},
  \bibinfo{author}{\bibfnamefont{F.}~\bibnamefont{{Machado}}},
  \bibinfo{author}{\bibfnamefont{N.~Y.} \bibnamefont{{Yao}}}, \bibnamefont{and}
  \bibinfo{author}{\bibfnamefont{M.~P.} \bibnamefont{{Zaletel}}},
  \bibinfo{journal}{arXiv e-prints} \bibinfo{eid}{arXiv:2110.00585}
  (\bibinfo{year}{2021}), \eprint{2110.00585}.

\bibitem[{\citenamefont{Lakshmanan}(2010)}]{Lakshmanan2010}
\bibinfo{author}{\bibfnamefont{M.}~\bibnamefont{Lakshmanan}},
  \bibinfo{journal}{Phil. Trans. R. Soc. A.} \textbf{\bibinfo{volume}{369}},
  \bibinfo{pages}{1280} (\bibinfo{year}{2010}).

\bibitem[{\citenamefont{Brown}(1963)}]{William1963}
\bibinfo{author}{\bibfnamefont{W.~F.} \bibnamefont{Brown}},
  \bibinfo{journal}{Phys. Rev.} \textbf{\bibinfo{volume}{130}},
  \bibinfo{pages}{1677} (\bibinfo{year}{1963}).

\bibitem[{\citenamefont{Lupo and Weber}(2019)}]{Lupo2019}
\bibinfo{author}{\bibfnamefont{C.}~\bibnamefont{Lupo}} \bibnamefont{and}
  \bibinfo{author}{\bibfnamefont{C.}~\bibnamefont{Weber}},
  \bibinfo{journal}{Phys. Rev. B} \textbf{\bibinfo{volume}{100}},
  \bibinfo{pages}{195431} (\bibinfo{year}{2019}).

\bibitem[{\citenamefont{Gambetta et~al.}(2019)\citenamefont{Gambetta, Carollo,
  Lazarides, Lesanovsky, and Garrahan}}]{Gambetta2019}
\bibinfo{author}{\bibfnamefont{F.~M.} \bibnamefont{Gambetta}},
  \bibinfo{author}{\bibfnamefont{F.}~\bibnamefont{Carollo}},
  \bibinfo{author}{\bibfnamefont{A.}~\bibnamefont{Lazarides}},
  \bibinfo{author}{\bibfnamefont{I.}~\bibnamefont{Lesanovsky}},
  \bibnamefont{and} \bibinfo{author}{\bibfnamefont{J.~P.}
  \bibnamefont{Garrahan}}, \bibinfo{journal}{Phys. Rev. E}
  \textbf{\bibinfo{volume}{100}}, \bibinfo{pages}{060105}
  (\bibinfo{year}{2019}).

\bibitem[{\citenamefont{Khasseh et~al.}(2019)\citenamefont{Khasseh, Fazio,
  Ruffo, and Russomanno}}]{Khasseh2019}
\bibinfo{author}{\bibfnamefont{R.}~\bibnamefont{Khasseh}},
  \bibinfo{author}{\bibfnamefont{R.}~\bibnamefont{Fazio}},
  \bibinfo{author}{\bibfnamefont{S.}~\bibnamefont{Ruffo}}, \bibnamefont{and}
  \bibinfo{author}{\bibfnamefont{A.}~\bibnamefont{Russomanno}},
  \bibinfo{journal}{Phys. Rev. Lett.} \textbf{\bibinfo{volume}{123}},
  \bibinfo{pages}{184301} (\bibinfo{year}{2019}).

\bibitem[{\citenamefont{Hurtado-Guti\'errez
  et~al.}(2020)\citenamefont{Hurtado-Guti\'errez, Carollo, P\'erez-Espigares,
  and Hurtado}}]{Hurtado2020}
\bibinfo{author}{\bibfnamefont{R.}~\bibnamefont{Hurtado-Guti\'errez}},
  \bibinfo{author}{\bibfnamefont{F.}~\bibnamefont{Carollo}},
  \bibinfo{author}{\bibfnamefont{C.}~\bibnamefont{P\'erez-Espigares}},
  \bibnamefont{and} \bibinfo{author}{\bibfnamefont{P.~I.}
  \bibnamefont{Hurtado}}, \bibinfo{journal}{Phys. Rev. Lett.}
  \textbf{\bibinfo{volume}{125}}, \bibinfo{pages}{160601}
  (\bibinfo{year}{2020}).

\bibitem[{\citenamefont{Ye et~al.}(2021)\citenamefont{Ye, Machado, and
  Yao}}]{Ye2021}
\bibinfo{author}{\bibfnamefont{B.}~\bibnamefont{Ye}},
  \bibinfo{author}{\bibfnamefont{F.}~\bibnamefont{Machado}}, \bibnamefont{and}
  \bibinfo{author}{\bibfnamefont{N.~Y.} \bibnamefont{Yao}},
  \bibinfo{journal}{Phys. Rev. Lett.} \textbf{\bibinfo{volume}{127}},
  \bibinfo{pages}{140603} (\bibinfo{year}{2021}).

\bibitem[{\citenamefont{Pizzi et~al.}(2021)\citenamefont{Pizzi, Nunnenkamp, and
  Knolle}}]{Pizzi2021}
\bibinfo{author}{\bibfnamefont{A.}~\bibnamefont{Pizzi}},
  \bibinfo{author}{\bibfnamefont{A.}~\bibnamefont{Nunnenkamp}},
  \bibnamefont{and} \bibinfo{author}{\bibfnamefont{J.}~\bibnamefont{Knolle}},
  \bibinfo{journal}{Phys. Rev. Lett.} \textbf{\bibinfo{volume}{127}},
  \bibinfo{pages}{140602} (\bibinfo{year}{2021}).

\bibitem[{\citenamefont{{Ament} et~al.}(2016)\citenamefont{{Ament},
  {Rangarajan}, {Parthasarathy}, and {Rakheja}}}]{Ament2016}
\bibinfo{author}{\bibfnamefont{S.}~\bibnamefont{{Ament}}},
  \bibinfo{author}{\bibfnamefont{N.}~\bibnamefont{{Rangarajan}}},
  \bibinfo{author}{\bibfnamefont{A.}~\bibnamefont{{Parthasarathy}}},
  \bibnamefont{and}
  \bibinfo{author}{\bibfnamefont{S.}~\bibnamefont{{Rakheja}}},
  \bibinfo{journal}{arXiv e-prints} \bibinfo{eid}{arXiv:1607.04596}
  (\bibinfo{year}{2016}), \eprint{1607.04596}.

\bibitem[{Sup()}]{Supplementary}
\bibinfo{howpublished}{See the Supplementary Material for the details of our
  numerical method, convergence check of our numerical result as well as an
  analysis of the finite size effect}.

\bibitem[{\citenamefont{Yang and Cai}(2021)}]{Yang2021}
\bibinfo{author}{\bibfnamefont{X.}~\bibnamefont{Yang}} \bibnamefont{and}
  \bibinfo{author}{\bibfnamefont{Z.}~\bibnamefont{Cai}},
  \bibinfo{journal}{Phys. Rev. Lett.} \textbf{\bibinfo{volume}{126}},
  \bibinfo{pages}{020602} (\bibinfo{year}{2021}).

\bibitem[{\citenamefont{{Chinzei} and {Ikeda}}(2021)}]{Chinzei2021}
\bibinfo{author}{\bibfnamefont{K.}~\bibnamefont{{Chinzei}}} \bibnamefont{and}
  \bibinfo{author}{\bibfnamefont{T.~N.} \bibnamefont{{Ikeda}}},
  \bibinfo{journal}{arXiv e-prints} \bibinfo{eid}{arXiv:2110.00591}
  (\bibinfo{year}{2021}), \eprint{2110.00591}.

\bibitem[{\citenamefont{Glauber}(1963)}]{Glauber1963}
\bibinfo{author}{\bibfnamefont{R.~J.} \bibnamefont{Glauber}},
  \bibinfo{journal}{J. Math. Phys.} \textbf{\bibinfo{volume}{4}},
  \bibinfo{pages}{294} (\bibinfo{year}{1963}).

\bibitem[{\citenamefont{Huse}(1986)}]{Huse1986}
\bibinfo{author}{\bibfnamefont{D.~A.} \bibnamefont{Huse}},
  \bibinfo{journal}{Phys. Rev. B} \textbf{\bibinfo{volume}{34}},
  \bibinfo{pages}{7845} (\bibinfo{year}{1986}).

\bibitem[{\citenamefont{Sandvik}(2010)}]{Sandvik2010}
\bibinfo{author}{\bibfnamefont{A.~W.} \bibnamefont{Sandvik}},
  \bibinfo{journal}{AIP Conference Proceedings}
  \textbf{\bibinfo{volume}{1297}}, \bibinfo{pages}{135} (\bibinfo{year}{2010}).

\bibitem[{\citenamefont{Magnus}(1954)}]{Magnus1954}
\bibinfo{author}{\bibfnamefont{W.}~\bibnamefont{Magnus}},
  \bibinfo{journal}{Communications on Pure and Applied Mathematics}
  \textbf{\bibinfo{volume}{7}}, \bibinfo{pages}{649} (\bibinfo{year}{1954}).

\bibitem[{\citenamefont{Korniss et~al.}(2000)\citenamefont{Korniss, White,
  Rikvold, and Novotny}}]{Korniss2000}
\bibinfo{author}{\bibfnamefont{G.}~\bibnamefont{Korniss}},
  \bibinfo{author}{\bibfnamefont{C.~J.} \bibnamefont{White}},
  \bibinfo{author}{\bibfnamefont{P.~A.} \bibnamefont{Rikvold}},
  \bibnamefont{and} \bibinfo{author}{\bibfnamefont{M.~A.}
  \bibnamefont{Novotny}}, \bibinfo{journal}{Phys. Rev. E}
  \textbf{\bibinfo{volume}{63}}, \bibinfo{pages}{016120}
  (\bibinfo{year}{2000}).

\bibitem[{\citenamefont{Fujisaka et~al.}(2001)\citenamefont{Fujisaka, Tutu, and
  Rikvold}}]{Fujisaka2001}
\bibinfo{author}{\bibfnamefont{H.}~\bibnamefont{Fujisaka}},
  \bibinfo{author}{\bibfnamefont{H.}~\bibnamefont{Tutu}}, \bibnamefont{and}
  \bibinfo{author}{\bibfnamefont{P.~A.} \bibnamefont{Rikvold}},
  \bibinfo{journal}{Phys. Rev. E} \textbf{\bibinfo{volume}{63}},
  \bibinfo{pages}{036109} (\bibinfo{year}{2001}).

\bibitem[{\citenamefont{Navarrete-Benlloch
  et~al.}(2017)\citenamefont{Navarrete-Benlloch, Weiss, Walter, and
  de~Valc\'arcel}}]{Navarrete2017}
\bibinfo{author}{\bibfnamefont{C.}~\bibnamefont{Navarrete-Benlloch}},
  \bibinfo{author}{\bibfnamefont{T.}~\bibnamefont{Weiss}},
  \bibinfo{author}{\bibfnamefont{S.}~\bibnamefont{Walter}}, \bibnamefont{and}
  \bibinfo{author}{\bibfnamefont{G.~J.} \bibnamefont{de~Valc\'arcel}},
  \bibinfo{journal}{Phys. Rev. Lett.} \textbf{\bibinfo{volume}{119}},
  \bibinfo{pages}{133601} (\bibinfo{year}{2017}).

\bibitem[{\citenamefont{Martin et~al.}(1973)\citenamefont{Martin, Siggia, and
  Rose}}]{Martin1973}
\bibinfo{author}{\bibfnamefont{P.~C.} \bibnamefont{Martin}},
  \bibinfo{author}{\bibfnamefont{E.~D.} \bibnamefont{Siggia}},
  \bibnamefont{and} \bibinfo{author}{\bibfnamefont{H.~A.} \bibnamefont{Rose}},
  \bibinfo{journal}{Phys. Rev. A} \textbf{\bibinfo{volume}{8}},
  \bibinfo{pages}{423} (\bibinfo{year}{1973}).

\bibitem[{\citenamefont{Kamenev}(2011)}]{Kamenev2011}
\bibinfo{author}{\bibfnamefont{A.}~\bibnamefont{Kamenev}},
  \emph{\bibinfo{title}{Field theory of non-equilibrium systems}}
  (\bibinfo{publisher}{~Cambridge University Press, Cambridge},
  \bibinfo{year}{2011}).

\bibitem[{\citenamefont{Altland and Simons}(2010)}]{Altland2010}
\bibinfo{author}{\bibfnamefont{A.}~\bibnamefont{Altland}} \bibnamefont{and}
  \bibinfo{author}{\bibfnamefont{B.}~\bibnamefont{Simons}},
  \emph{\bibinfo{title}{Condensed matter field theory}}
  (\bibinfo{publisher}{Cambridge University Press, Cambridge},
  \bibinfo{year}{2010}).

\bibitem[{\citenamefont{Natsheh et~al.}(2021)\citenamefont{Natsheh, Gambassi,
  and Mitra}}]{Natsheh2021}
\bibinfo{author}{\bibfnamefont{M.}~\bibnamefont{Natsheh}},
  \bibinfo{author}{\bibfnamefont{A.}~\bibnamefont{Gambassi}}, \bibnamefont{and}
  \bibinfo{author}{\bibfnamefont{A.}~\bibnamefont{Mitra}},
  \bibinfo{journal}{Phys. Rev. B} \textbf{\bibinfo{volume}{103}},
  \bibinfo{pages}{224311} (\bibinfo{year}{2021}).

\end{thebibliography}

\newpage


\begin{center}
\textbf{\Large{Supplemental material}}
\end{center}

In this supplementary material, we first provide some details of the Heun algorithm used in our simulation, then we numerically check the convergence of our results with the finite discrete time step $\Delta t$. Finally, we discuss the dependence of our results on the initial states as well as the finite size effect in our simulations.

\section{Solving the stochastic Landau-Lifshitz-Gilbert equation: Heun algorithm}
In this section, we first derive the Stratonovich  form of the stochastic LLG equation, then formulate the Heun algorithm for Stratonovich Stochastic Differential equations (SDE). A stochastic LLG equation reads:
\begin{equation}
\dot{\mathbf{s}}_i=\mathbf{h}_i\times \mathbf{s}_i- \lambda  \mathbf{s}_i\times (\mathbf{s}_i\times \mathbf{h}_i) \label{eq:EOM}
\end{equation}
where $\mathbf{s}_i$ is a unit vector.  $\mathbf{h}_i(t)=\mathbf{h}^0_i(t)+\bm{h}^T_i(t)$ is the effective magnetic field $\mathbf{h}^0_i(t)=[V(t) \bar{s}_i^x, 0, h_z]$ with $\bar{s}_i^x= \frac 12\sum_{\langle j\rangle} s_j^x$ where the summation is over the nearest neighboring sites of site i.  $\bm{h}^T_i(t)$ is a three-dimensional(3D) stochastic magnetic field representing the thermal noise satisfying:
\begin{equation}
\langle h_i^{T\alpha}(t)h_j^{T\beta}(t')\rangle_{\bm\xi}=\mathcal{D}^2\delta_{\alpha\beta}\delta_{ij} \delta(t-t') \label{eq:noise}
\end{equation}
where $\alpha,\beta=x,y,z$ and $\mathcal{D}$ is the strength of the noise satisfying $\mathcal{D}^2=2T \lambda$, and the average $\langle\rangle_{\bm\xi}$ is over all the trajectories of noises.

To simulate the thermal noise numerically, we discretize the time with the time step $\Delta t$ of the numerical method.  Provided that the spin configuration in the $m$-th time step ($t_m=m\Delta t$) is defined as $\{\mathbf{s}_i^{m}\}$,  in the Heun algorithm,  the calculation of   $\{\mathbf{s}_i^{m+1}\}$ can be divided into two steps. We first calculate:
\begin{equation}
\tilde{\mathbf{s}}_i^{m+1}=\mathbf{s}_i^{m}+[\mathbf{h}^{m}_i\times \mathbf{s}_i^m- \lambda  \mathbf{s}_i^{m}\times (\mathbf{s}_i^{m}\times \mathbf{h}^{m}_i)]\Delta t \label{eq:mid}
\end{equation}
with $\mathbf{h}_i^m=\mathbf{h}^0_{i,m}+\bm{\tilde{h}}^T_{i,m}$, where $\mathbf{h}^0_{i,m}=\mathbf{h}^0_i(t_m)$ and $\bm{\tilde{h}}^T_{i,m}$ is a stochastic magnetic field satisfying:
\begin{equation}
\tilde{h}^{T\alpha}_{i,m}=\frac{\mathcal{D}}{\sqrt{\Delta t}}\xi_{i,m}^\alpha
\end{equation}
where $\xi_i^\alpha$ is a random number satisfying the Gaussian distribution with $\mathcal{N}(0,1)$: $\langle \xi_i^\alpha\rangle_{\bm\xi}=0$, $\langle (\xi_i^\alpha)^2\rangle_{\bm\xi}=1$.

In the Heun algorithm, $\mathbf{s}_i$ at the $m+1$-th time step can be expressed as:
\begin{eqnarray}
\nonumber &\mathbf{s}_i^{m+1}&=\mathbf{s}_i^{m}+\frac{\Delta t}2\{ \mathbf{h}^{m}_i\times \mathbf{s}_i^m- \lambda  \mathbf{s}_i^{m}\times (\mathbf{s}_i^{m}\times \mathbf{h}^{m}_i)\\
&+&\tilde{\mathbf{h}}^{m+1}_i\times \tilde{\mathbf{s}}_i^{m+1}- \lambda  \tilde{\mathbf{s}}_i^{m+1}\times (\tilde{\mathbf{s}}_i^{m+1}\times \tilde{\mathbf{h}}^{m+1}_i)\}
\end{eqnarray}
where $\tilde{\mathbf{s}}_i^{m+1}$ has been defined in Eq.(\ref{eq:mid}), and $\tilde{\mathbf{h}}_i^{m+1}=\mathbf{h}^0_{i,m+1}+\bm{\tilde{h}}^T_{i,m}$.

 \begin{figure}[htb]
\includegraphics[width=0.99\linewidth,bb=110 57 810 550]{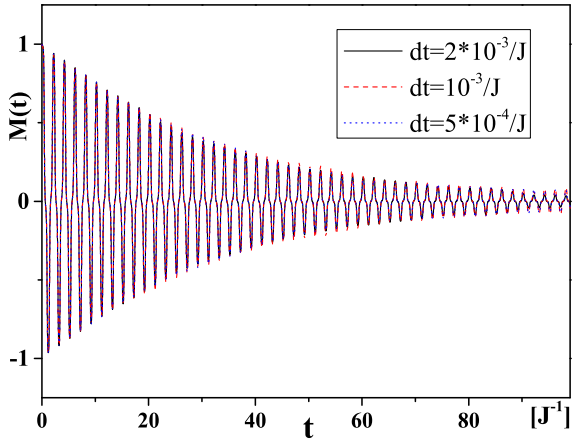}
\caption{(Color online) Comparison of the M(t) with parameters $L=8$, $\mathcal{D}=0.2782$, $\mathcal{N}=1000$ and  different $\Delta t$,
} \label{fig:SM1}
\end{figure}

\begin{figure}[htb]
\includegraphics[width=0.99\linewidth,bb=235 48 1400 559 ]{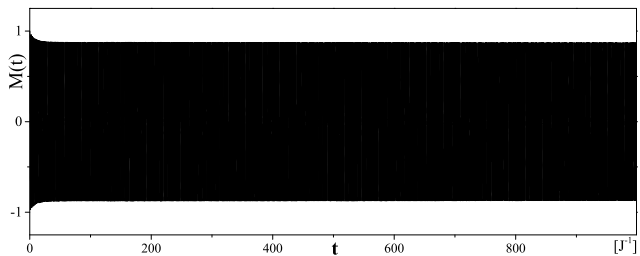}
\caption{The long-time dynamics of the FM order parameter M(t) in the DTC phase with parameters $L=20$, $D=0.26J$.} \label{fig:SM1b}
\end{figure}

 \begin{figure}[htb]
\includegraphics[width=0.99\linewidth,bb=110 57 810 550]{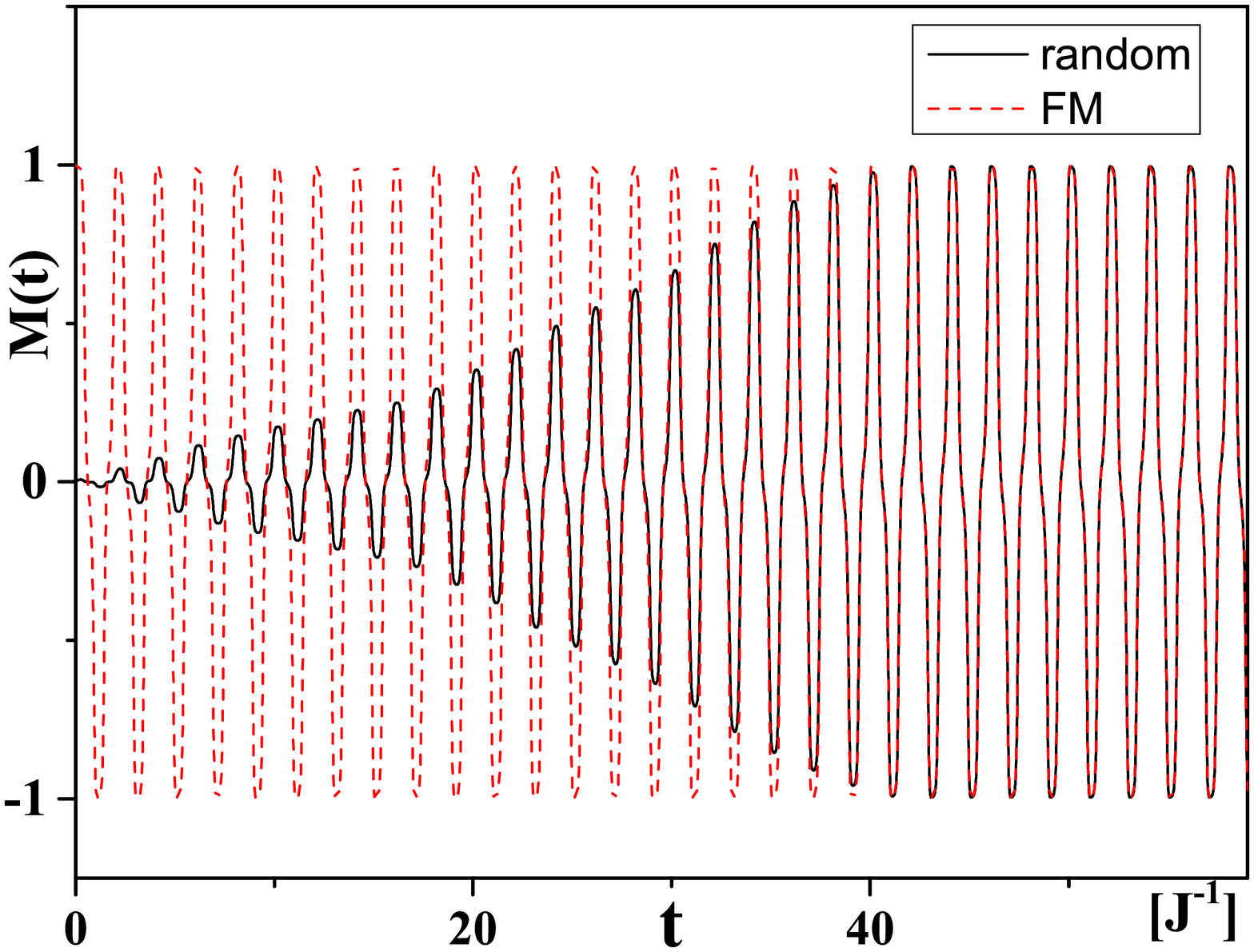}
\caption{(Color online) Comparison of the M(t) starting from different initial states with parameters $L=20$, $\mathcal{D}=0$.
} \label{fig:SM2}
\end{figure}

 \begin{figure}[htb]
\includegraphics[width=0.99\linewidth,bb=110 57 910 550]{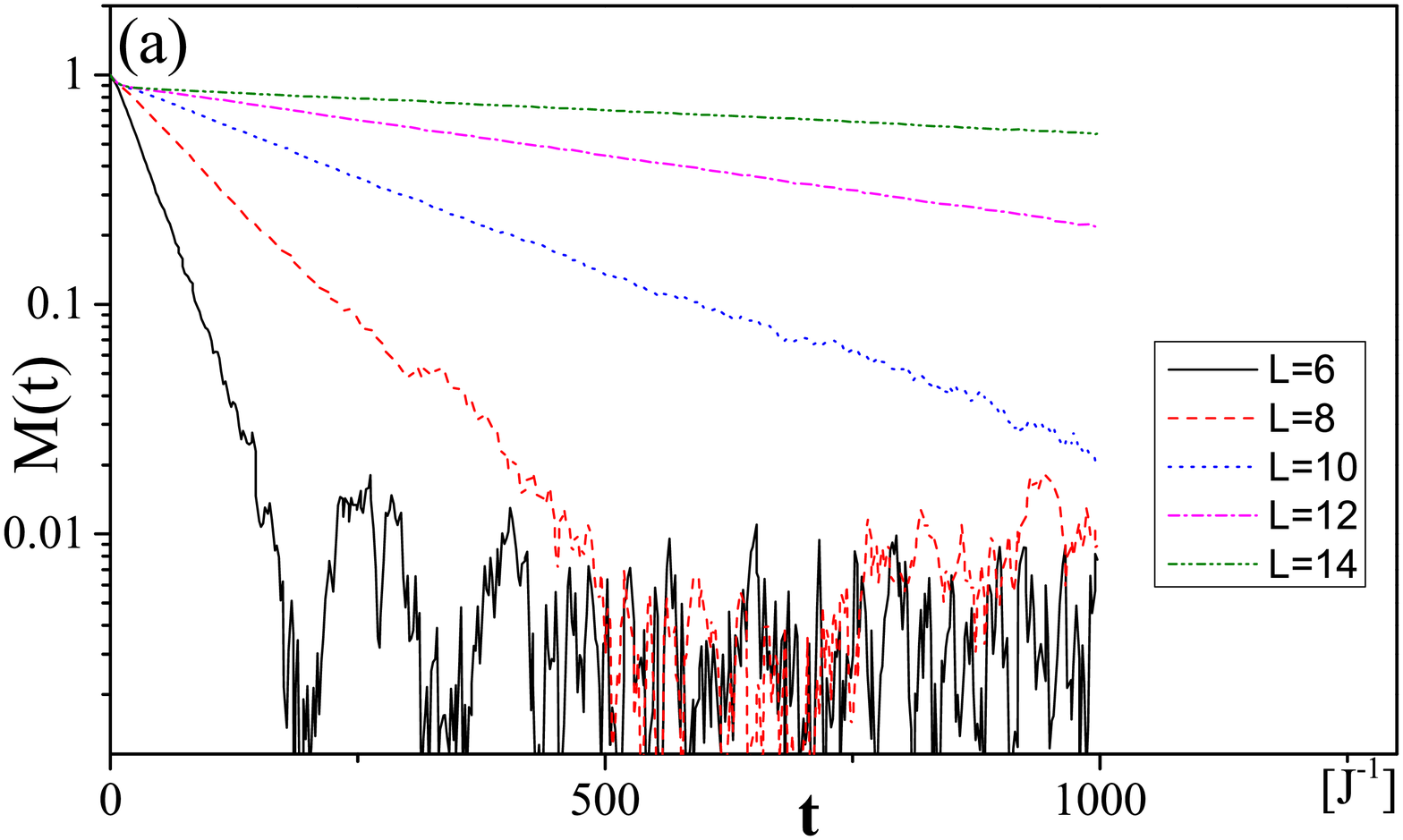}
\includegraphics[width=0.99\linewidth,bb=110 57 910 550]{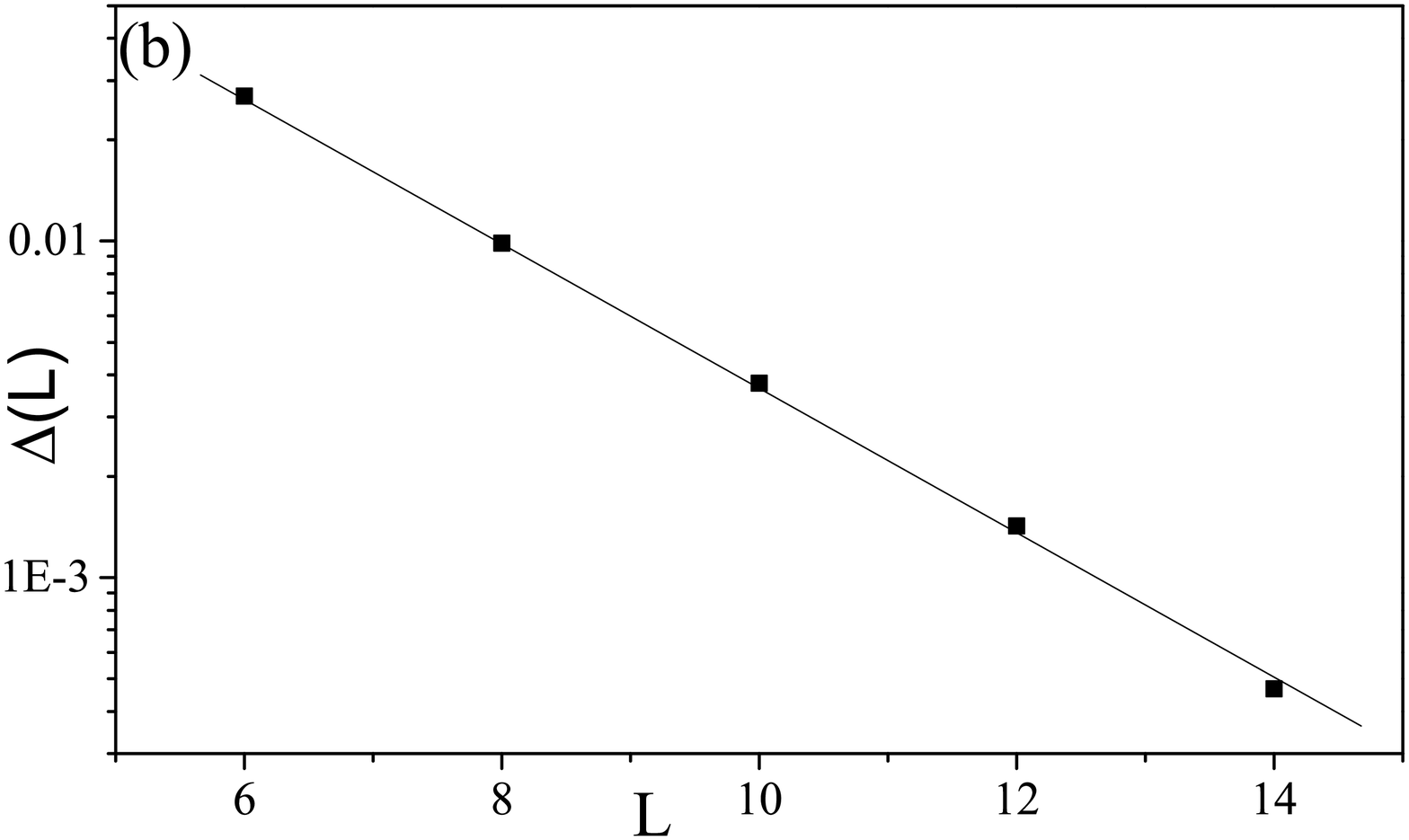}
\caption{(Color online) (a) The envelop of M(t) in the DTC phases of small systems with various system sizes and parameter $D=0.26J<D_c$. (b) The decay rate $\Delta(L)$ as a function of system size L.
} \label{fig:SM3}
\end{figure}

\begin{figure}[htb]
\includegraphics[width=0.99\linewidth,bb=235 48 1200 559 ]{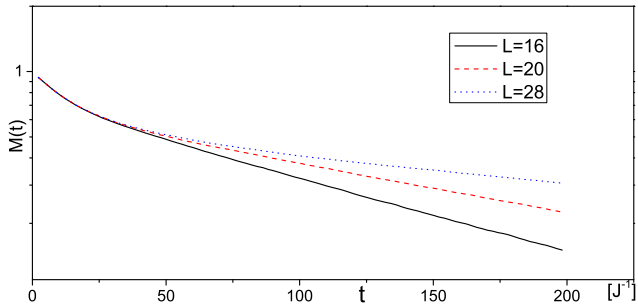}
\caption{Replotting Fig. 2 (d) in the main text using a semi-log plot} \label{fig:SM4}
\end{figure}

\section{ Convergence of numerical results}
In our simulation, we choose the time step $\Delta t=10^{-3}J^{-1}$. In general, for  stochastic differential equations,  the dependence of the numerical results on $\Delta t$ is more subtle than the deterministic ones (as shown in Eq.(\ref{eq:noise}), the strength of the stochastic magnetic fields depend on $\Delta t$) thus we need to carefully examine the convergence of our result (especially the power-law decays) with $\Delta t$, and preclude the possibility that it is an artifact because of the finite $\Delta t$ we choose. To this end, we choose different $\Delta t=2\times 10^{-3}, 10^{-3}$ and $5\times10^{-4}$, and compare their results. As shown in Fig.\ref{fig:SM1}, the results with  $\Delta t= 10^{-3}$ and $5\times10^{-4}$ agree with each other very well within the statistical error bar from the ensemble average of the  trajectories of noise, which indicates that the $\Delta t$ we choose in our simulation is sufficiently small that allows us to ignore the finite-$\Delta t$ induced errors.

The typical time scale of the simulation in the maintext is up to $\sim 100J^{-1}$ (50 DTC periods).  One may wonder whether such a time scale is in a prethermal regime, and longer simulation may give rise to qualitatively different dynamics. To verify this point, we extended the simulation time up to $10^3 J^{-1}$ (500 periods of the DTC) , which is the maximum accessible time scale in our numerical simulations considering the accumulated errors in the Heun algorithm.  As shown in Fig.\ref{fig:SM1b}, the DTC order parameter barely decays for a sufficiently large system ($L=20$). However,  as we will show below,  for a small system, there is an exponential decay even at temperature below the critical temperature, which can be considered as a finite-size effect.

\section{Dependence on the initial state}
In our simulation in the main text, we start from an spatially homogeneous initial state (the ground state of the system Hamiltonian at $t=0$), where the FM order parameter $M(t)$ is proportional to the auto-correlation function in time $C(t)$. It is important to check that our results doesn't crucially depend on this specific choice of the initial state. To this end, we choose an inhomogeneous random initial state (paramagnetic state with $M(t=0)=0$): for each site, we choose the the initial spin as $\bm{s}_i^0=[s_i^x,0,s_i^z]$, where $s_i^x$ is an random number different from site by site and uniformly distributed within [-1,1], the corresponding z-component spin is chosen as $s_i^z=\sqrt{1-[s_i^x]^2}$.  We compare the $M(t)$ from an uniform ferromagnetic (FM) and random initial states.  As shown in Fig.\ref{fig:SM2}, after a relaxation dynamics, the time evolution of $M(t)$ in these two cases agree with each other very well within the statistical error, indicating a rapid loss of the memory of the initial state information, which can be understood as a consequence of the coupling to the thermal bath.

\section{Finite size effect in the DTC phases and critical point}
All the numerical simulations in the main text are preformed in finite size systems. In equilibrium physics, it is well known that the spontaneous symmetry breaking can only occur in thermodynamic limit, while a symmetry breaking phase in a finite system has a life time, which exponentially diverges with the system size. Since the DTC phases in our model also spontaneously breaks the discrete translational symmetry, it is interesting to ask whether such a non-equilibrium symmetry breaking phase possesses similar properties.

To this end, we study the dynamics of the FM order parameter for small systems with  fixed $D=0.26J$ below the critical temperature. As shown in Fig.\ref{fig:SM3} (a), M(t) in small systems with various system size indeed decay exponentially in time ($|M(t)|\sim e^{-\Delta(L) t})$, which indicates a finite life time $\tau_c\sim \Delta(L)^{-1}$ for the DTC phase. A finite size scaling of $\Delta(L)$ is plotted in Fig.\ref{fig:SM3} (b), which shows an exponential decay of $\Delta(L)$ with system size (an exponential divergence of $\tau_c$). This result indicates that despite the genuine non-equilibrium feature of the DTC phases, it share with some common features with the spontaneous symmetry breaking phases in equilibrium systems.

Now we focus on the finite-size effect at the critical point. As shown in Fig.\ref{fig:SM4}, for a finite system at the critical point, the long-time dynamics also exhibit an exponential decay due to the finite-size effect, whose exponent $\Delta$ depends on the system size. However, different from the DTC phase below the critical temperature, at the critical point, $\Delta(L)$ decay algebraically instead of exponentially with $L$ (see the inset of Fig.2 (d) in the maintext), from which we can  extract a dynamical critical exponent $z$ ($\Delta(L)\sim L^{-z}$).

\end{document}